\documentclass[twocolumns]{aa} 

\usepackage{graphicx}
\usepackage{epstopdf}
\usepackage{deluxetable}
\usepackage{times,epsfig} 
\usepackage{mathrsfs}  
\usepackage{natbib}
\usepackage{amssymb}
\bibpunct{(}{)}{;}{a}{}{,} 
\usepackage[english]{babel}
\usepackage{longtable}
\usepackage[english]{babel}
\usepackage{caption}
\usepackage{url}
\usepackage{hyperref}

\authorrunning{F. Taddia, et al.}
\titlerunning{PTF11mnb: the first analog of SN~2005bf}

\begin{document}

\title{PTF11mnb: the first analog of supernova 2005bf}
\subtitle{A long-rising, double-peaked supernova Ic from a massive progenitor}

\author{F. Taddia\inst{1}, 
 J. Sollerman\inst{1}, 
 C. Fremling\inst{1,5},
 E. Karamehmetoglu\inst{1},
 R.~M. Quimby\inst{2,3},
 A. Gal-Yam\inst{4},
 O. Yaron\inst{4},
 M.~M. Kasliwal\inst{5},
 S.~R. Kulkarni\inst{5},
 P.~E. Nugent\inst{6,7},
  G. Smadja\inst{8},
  C. Tao\inst{9,10}
}
\institute{The Oskar Klein Centre, Department of Astronomy, Stockholm University, AlbaNova, 10691 Stockholm, Sweden\\
 \email{francesco.taddia@astro.su.se}
\and Department of Astronomy, San Diego State University, San Diego, CA 92182, USA
\and Kavli IPMU (WPI), UTIAS, The University of Tokyo, Kashiwa, Chiba 277-8583, Japan
\and Benoziyo Center for Astrophysics, Weizmann Institute of Science, Rehovot 76100, Israel
\and Cahill Center for Astrophysics, California Institute of Technology, Pasadena, CA 91125, USA
\and Astronomy Department, University of California at Berkeley, Berkeley, CA 94720, USA
\and Lawrence Berkeley National Laboratory, 1 Cyclotron Road, MS 50B-4206, Berkeley, CA 94720, USA
\and Universit\'e de Lyon 1, Villeurbanne; CNRS/IN2P3, Institut de Physique Nucl\'eaire de Lyon, F-69622, Lyon, France
\and Tsinghua Center for Astrophysics, Tsinghua University, Beijing 100084, China
\and Centre de Physique des Particules de Marseille, Aix-Marseille Universit\'e, CNRS/IN2P3, 163 avenue de Luminy - Case 902 - 13288 Marseille Cedex 09, France
}

\date{Received; accepted}

\abstract
{}
{We study PTF11mnb, a He-poor supernova (SN) whose light curves resemble those of
  SN~2005bf, a peculiar double-peaked stripped-envelope (SE)~SN, until the declining phase after the main peak. 
We investigate the mechanism powering its light curve and the nature of its progenitor star.}
{Optical photometry and spectroscopy of PTF11mnb are presented. Light
  curves, colors and spectral properties are compared to those of
  SN~2005bf and normal SE~SNe. A bolometric light curve is built and
  modeled with the SNEC hydrodynamical code explosion of a MESA
  progenitor star, as well as with semi-analytic models.}
{The light curve of PTF11mnb turns out to be similar to that of 
  SN~2005bf until $\sim$50~d, when the main (secondary) peaks occur at
  $-18.5$~mag. The early peak occurs at $\sim$20~d, and is about 1.0
  mag fainter. After the main peak, the decline rate of PTF11mnb is 
  remarkably slower than what was observed in SN~2005bf, and it traces well 
  the $^{56}$Co decay rate. The spectra of PTF11mnb reveal a SN~Ic, with no 
  traces of He unlike in the case of SN~Ib 2005bf, although with velocities comparable to those of SN~2005bf. The whole evolution of the bolometric light curve is well reproduced by the explosion of a massive
  ($M_{ej}~=$~7.8~$M_{\odot}$), He-poor star characterized by a double-peaked
  $^{56}$Ni distribution, a total $^{56}$Ni mass of 0.59~$M_{\odot}$ and an
  explosion energy of 2.2$\times$10$^{51}$ erg. Alternatively, a normal SN~Ib/c explosion
(M($^{56}$Ni)$=$0.11~$M_{\odot}$, $E_{K}$~=~0.2$\times$10$^{51}$~erg,
$M_{ej}~=$~1~$M_{\odot}$) can power the first peak while a magnetar
(with a magnetic field characterized by $B$=5.0$\times$10$^{14}$~G,
and a rotation period of $P=18.1$~ms) provides energy for the
main peak. The early $g$-band light curve can be fit with a
shock-breakout cooling-tail or an extended envelope model,
from which a radius of at least 30 $R_{\odot}$ is obtained.}
{We presented a scenario where PTF11mnb was the explosion of a massive, He-poor star, characterized by a double-peaked $^{56}$Ni
  distribution. In this case, the ejecta mass and the absence of He imply a large ZAMS mass ($\sim85~M_{\odot}$) for the progenitor, which most likely was a Wolf-Rayet star, surrounded by an extended envelope formed either by a pre-SN eruption or due to a binary configuration. Alternatively, PTF11mnb could be powered by a SE~SN with a less massive progenitor during the first peak and by a magnetar afterwards.}


\maketitle
\section{Introduction}
\label{sec:intro}
The majority of stripped-envelope (SE) supernovae (SNe) share rather similar light-curve properties \citep[see e.g.,][]{drout11,cano13,bianco14,taddia15,lyman16,prentice16,taddia17}, with typical rise times of $15-25$ days in the optical, slower decline and peak magnitudes between $-$17 and $-$18 mag. Most of these SNe are inferred to have ejected $2-4~M_{\odot}$,
with energies of a few 10$^{51}$~erg and $^{56}$Ni masses between
0.15--0.20~$M_{\odot}$. Assuming these SNe gave birth to compact remnants with the mass of a neutron star, their rather modest ejecta masses suggest that
SE SNe arise from relatively low-mass
(M$_{ZAMS}~\lesssim$~15$ M_{\odot}$) progenitors, which then must have
been stripped of their H/He envelopes 
by a companion star \citep[e.g.,][]{yoon10,eldridge13,lyman16}.

A few SE~SN events show clearly different properties, with light
curves characterized by a more complex morphology and different time
scales. Two examples are SN~2011bm \citep{valenti12} and iPTF15dtg
\citep{taddia16} which are characterized by longer rise-times
($\sim$30--40~d) and broader light curves. iPTF15dtg (and perhaps SN~2011bm)
also exhibits an early declining phase in the optical light curves \citep{taddia16}. 
These SNe may have large ejecta masses ($\sim$10~$M_{\odot}$) compared to
normal SE~SNe, 
and might come from single massive stars.
Another unusual SE~SN is the double-peaked SN~2005bf. This transient was discovered by
\citet{monard05} and became one of the most peculiar SE~SNe ever
observed.  SN~2005bf has remained unique until the discovery of
PTF11mnb, which we present in this paper. 

Therefore, we start by reviewing the main properties of SN~2005bf as presented in the literature.

\citet{anupama05} reported optical photometry and spectroscopy around
and post peak for SN~2005bf. This SN revealed a Type Ib (helium rich)
spectrum at peak, which occurred unusually late for a SE~SN
($\sim$40~d). 
The peak bolometric magnitude of $-$18 mag was consistent with those
of luminous SNe~Ibc. \citet{anupama05} observed that the \ion{He}{i}
lines were less blueshifted than the \ion{Fe}{} lines, and reported
traces of \ion{H}{} lines at 15000 km~s$^{-1}$. These authors suggested that SN~2005bf was the explosion of a massive \ion{He}{} star with some \ion{H}{} left.

\citet{tominaga05} reported that the light curve of SN~2005bf actually
had two maxima, the first at $\sim$20~d and the main (secondary) peak
at 40~d. After the second peak the light curve faded
rapidly. \citet{tominaga05} noted that the \ion{He}{I} lines 
strengthened and their velocities increased with time. 
A double-peaked $^{56}$Ni distribution was proposed to explain the two
maxima, with a small amount at high
velocity and most of the $^{56}$Ni  at low velocity. 
The fast decline after peak could be due to $\gamma$-rays escaping from low-density regions.
The evolution of the \ion{He}{} lines was then explained by enhanced
$\gamma$-ray deposition in the \ion{He}{} layer with time, as these
$\gamma$-rays were leaking out from the core. From their models,
\citet{tominaga05} estimated a large ejecta mass ($\sim$6--7
$M_{\odot}$), kinetic energy of $(1.0-1.5)\times10^{51}$~ergs, and a large
$^{56}$Ni mass ($\sim$0.32~$M_{\odot}$). The progenitor was claimed to be a WN star, whose double-peaked $^{56}$Ni distribution could possibly be due to jets that did not reach the \ion{He}{} layer.

\citet{folatelli06} 
presented observations of SN~2005bf covering the first
$\sim$100 days after discovery. Their spectroscopic observations
revealed that SN~2005bf exhibited increasingly stronger He lines. Furthermore, high velocity absorption lines were observed during the initial peak along with lower velocity line components (see also \citealp{parrent07}).
The scenario favored by \citet{folatelli06} was an energetic and
asymmetric explosion 
of a massive (8.3~$M_{\odot}$) WN
star almost completely stripped of its H envelope. 
\citet{folatelli06} attributes the high velocity features, the early spectrum, and the existence of the first peak, to a polar
explosion containing only part of the total mass; that early explosion
was then followed by the explosion of the rest of the star, which produced the main peak and the helium-rich spectra.

\citet{maeda07} presents nebular spectra ($\sim$300~d) and late-time
photometry of SN~2005bf. 
The emission line analysis reveals 
a blueshift compatible with a blob or a unipolar jet 
(or self- absorption within the ejecta) containing only tenths of a
solar mass of ejecta and a small amount (0.02$-$0.06~$M_{\odot}$) of $^{56}$Ni.
The late-time photometry sets an upper limit of 0.08~$M_{\odot}$ for
the $^{56}$Ni mass, 
in apparent contradiction to the high value 
derived from the main peak. To explain this discrepancy, and the fast decline of the light curve after peak, \citet{maeda07} suggest an alternative 
scenario where the powering source of SN~2005bf is a magnetar. The strong asymmetry of the explosion of SN~2005bf was also confirmed
by the spectropolarimetric observations by \citet{maund07} and \citet{tanaka09}. 

In summary, SN~2005bf showed the following peculiarities: 
a) a unique double peak in the light curve, with the first maximum
occurring at the same phase ($\sim$20 d) and with the same absolute magnitude of
a regular SN Ibc, while the main peak occurred relatively late
($\sim$40~d); b) a fast decline post peak and a very low optical luminosity at late epochs; c) increasingly stronger and faster He lines; d) multiple velocity components for some of the lines. 

In this paper, we present PTF11mnb, the first SN~2005bf-like event. With ``SN~2005bf-like" we mean a SE~SN with a first peak similar to that of a normal SN Ib/c, followed by a brighter peak occurring on a longer time scale. 
PTF11mnb was a He-poor (Type~Ic) SN whose pre-main peak optical light curves closely resemble those of
SN~2005bf. Both SNe show a double peak at similar phases and absolute
magnitudes, but PTF11mnb declines slower after the main maximum. 
Furthermore, PTF11mnb never shows \ion{He}{} in the spectrum and thus never becomes a Type Ib.  
For SN~PTF11mnb, we suggest a scenario where the progenitor was a massive single star with a double-peaked $^{56}$Ni distribution powering the SN rather than a magnetar, even though we cannot exclude the presence of a central engine.

The structure of the paper is as follows: In Sect.~\ref{sec:obs} we describe the discovery, observations and data reduction; in Sect.~\ref{sec:hg} we present the host galaxy. Section~\ref{sec:lc} includes the analysis of the SN light curves, whereas Sect.~\ref{sec:spec} includes that of the SN spectra. In Sect.~\ref{sec:model} we build and model the bolometric light curve of PTF11mnb. The main results are discussed in Sect.~\ref{sec:discussion}, whereas our conclusions are given in Sect.~\ref{sec:conclusions}.

\section{Observations and data reduction}
\label{sec:obs}

The Palomar Transient Factory \citep{rau09,law09} first detected
PTF11mnb at
RA~$=$~00:34:13.25 and DEC~$=$~+02:48:31.4 (J2000.0) on JD~2455804.857
(Aug. 31 2011)
($g~=~21.07\pm$0.31~mag ) using the 48-inch Samuel Oschin telescope
(P48) at Palomar Observatory, equipped with the 96~Mpixel mosaic
camera CFH12K \citep{rahmer08}. 
The SN was not detected on JD~2455803.824 (i.e., 1.033 days before
first detection) at limiting magnitude 
$g~\geq21.41$~mag. In the following, we adopt the average between the
epochs of last non-detection and discovery as the explosion date
($t_{\rm explo}=$JD~2455804.341$\pm$0.516). 
Throughout the paper we express the phase in SN rest-frame days since
explosion.  

PTF11mnb was observed with P48 in the $g$ band until $\sim$60~d and
until $\sim$50 days in the $r$ band. The P48 photometry was reduced with the FPipe pipeline presented by \citet{fremling16}, which gives very similar results to those obtained by using the standard  Palomar Transient Factory Image Differencing and Extraction (PTFIDE) pipeline \citep{masci17}.  
FPipe subtracts the host-galaxy template and performs point-spread-function (PSF) photometry on each SN image.

\begin{figure}
\centering
\includegraphics[width=9cm]{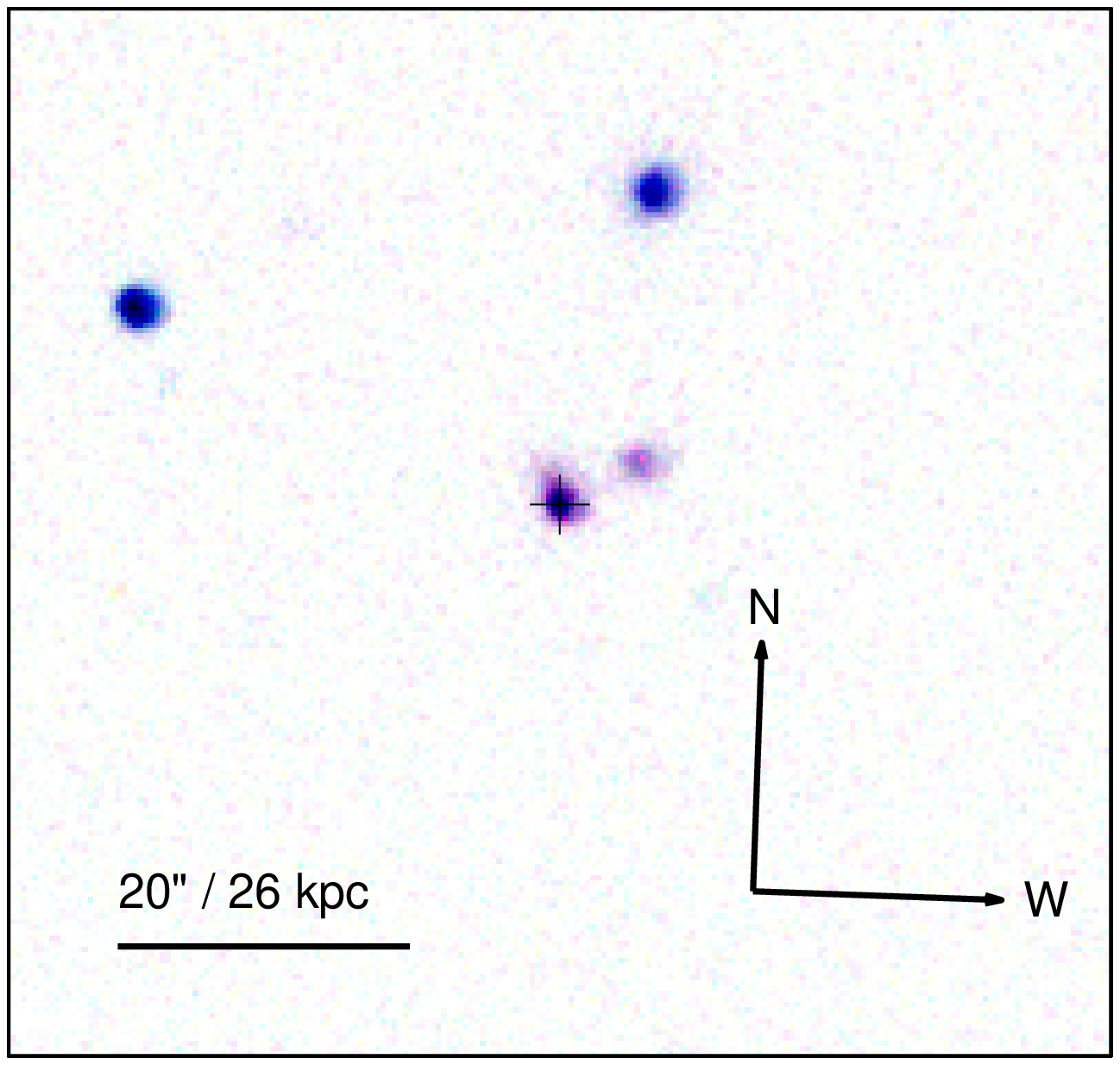}\\
\includegraphics[width=9cm]{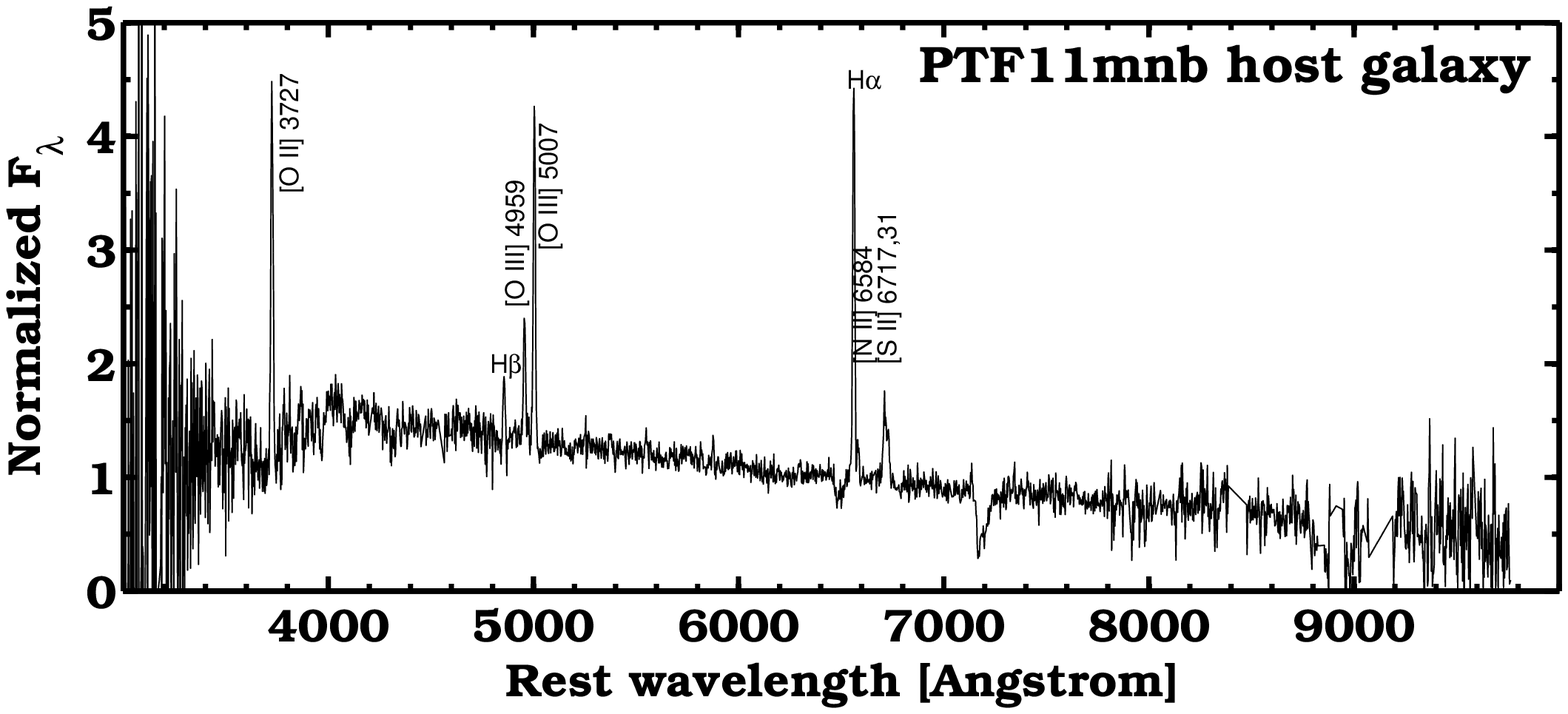}
\caption{\label{fc} (Top panel) PTF11mnb (marked by a black cross) and its host galaxy in a composite $Bgr$ image (with inverted colors) from P60 observations taken on Oct. 28 2011. The orientation of the image is indicated in the bottom-right corner, whereas the scale is shown in the bottom-left. (Bottom panel) TNG spectrum of the host galaxy of PTF11mnb, obtained in 2016. The main emission lines are identified.}
\end{figure}

We also made use of the Palomar 60-inch telescope (P60; \citealp{cenko06}), with which we observed the SN in $Bgri$ bands, starting at $\sim$20~d with the $r$ band ($Bgi-$band coverage started from $\sim$40~d). The SN was detected for the last time at $\sim$140 days with P60. 
A P60 composite $Bgr-$band image of PTF11mnb and its host galaxy from 28 Oct. 2011 is shown in the top panel of Fig.~\ref{fc}. 
The P60 images were also reduced with FPipe, making use of Sloan Digital Sky Survey (SDSS; \citealp{ahn14}) images as templates. We used the SDSS stars in the SN field to calibrate the P60 photometry. 
The final light curves are presented after combining magnitudes
obtained on the same night. We summarize all the photometric observations in Table~\ref{tab:phot}. 

We present five optical spectra covering the epochs between one
and five months after explosion. These spectra were obtained with  four different telescopes: 
Keck I $+$ the Low Resolution Imaging Spectrometer (LRIS, \citealp{oke95}), the Palomar 200-inch Hale Telescope (P200) $+$ the Double Spectrograph (DBSP, \citealp{oke82}), the Kitt Peak National Observatory 4-meter telescope (KPNO4m) $+$ the Ritchey-Chretien Spectrograph (RC spec), and the University of Hawai'i 88-inch (2.2-meter) telescope
(UH88) $+$ the Supernova Integrated Field Spectrograph (SNIFS). 
Spectral reductions were carried out with all the standard procedures, including wavelength calibration via an arc lamp, and flux calibration with a spectrophotometric standard star. 
In 2016 we also obtained a host-galaxy spectrum with the Telescopio Nazionale Galileo (TNG) $+$ the Device Optimized for the LOw RESolution (DOLORES). 
This host-galaxy spectrum is shown in the bottom panel of Fig.~\ref{fc}. Table~\ref{tab:spectra} shows our spectral log.

\section{Host galaxy}
\label{sec:hg}

The host galaxy of PTF11mnb is named SDSS J003413.34$+$024832.9.
We determined its redshift to be $z~=~$0.0603$\pm$0.0001  
from the
Gaussian fits of the H$\alpha$ and [\ion{O}{iii}]~$\lambda$5007
host-galaxy emission lines superimposed on the SN spectra (see
Sect.~\ref{sec:spec}), as well as from the host-galaxy spectrum obtained in 2016 (see Fig.~\ref{fc}). 
This redshift corresponds to a
luminosity distance $D_L$~$=$~268.5~Mpc and distance modulus
$\mu~=$37.14~mag when WMAP 5-years cosmological parameters
\citep{komatsu09} are assumed. 

Given the absence of any narrow \ion{Na}{i}~D absorption line at the host-galaxy redshift, we assume that no host extinction affected PTF11mnb. The Galactic extinction in the $Bgri$ bands is $A_B~=$~0.067~mag, $A_g~=$~0.061~mag, $A_r~=$~0.042~mag, and $A_i~=$~0.031~mag (\citealp{sf11}).

SDSS J003413.34$+$024832.9 has integrated magnitudes of  
$M_g~=~-17.8$~mag, $M_r~=~-18.0$~mag, $M_i~=~-18.3$~mag. 
Based on these absolute magnitudes, a global metallicity of 12$+$log(O/H)~$=$~8.29 ($Z/Z_{\odot}~=~0.4$, where $Z_{\odot}=8.69$, \citealp{asplund09}) is inferred from the luminosity-color-metallicity relation by \citet{sanders13}, or $Z/Z_{\odot}=0.25$ if we make use of the luminosity-metallicity relation by \citet{arcavi10}.

Since PTF11mnb sits on a bright \ion{H}{ii} region, we could measure
the host emission-line fluxes at the exact SN position from its
spectra. Using the high signal-to-noise Keck spectrum, we derived the
line ratios needed for the O3N2 method by \citet{pp04}, which resulted
in a metallicity of 12$+$log(O/H)~$=$~8.29$\pm$0.20
($Z/Z_{\odot}~=~0.40$). This value is lower than that for most of the normal SNe~Ic \citep[e.g.,][]{sanders12}, and more similar to
that of the explosion site of the peculiar SN~Ic iPTF15dtg \citep{taddia16}. 

We also notice that the host galaxy of SN~2005bf shows larger metal abundances.
From a Sloan Digital Sky Survey (SDSS) spectrum of an \ion{H}{ii} region in a spiral arm of
SN~2005bf's host, only marginally closer to the host center than
SN~2005bf itself, we measured an O3N2 oxygen abundance of 8.76$\pm$0.14, 
which is about solar \citep{asplund09}. 

We summarize the main host-galaxy properties in Table~\ref{tab:propgal}.

\section{Light curves}
\label{sec:lc}

\begin{figure*}
\centering
\includegraphics[width=18cm]{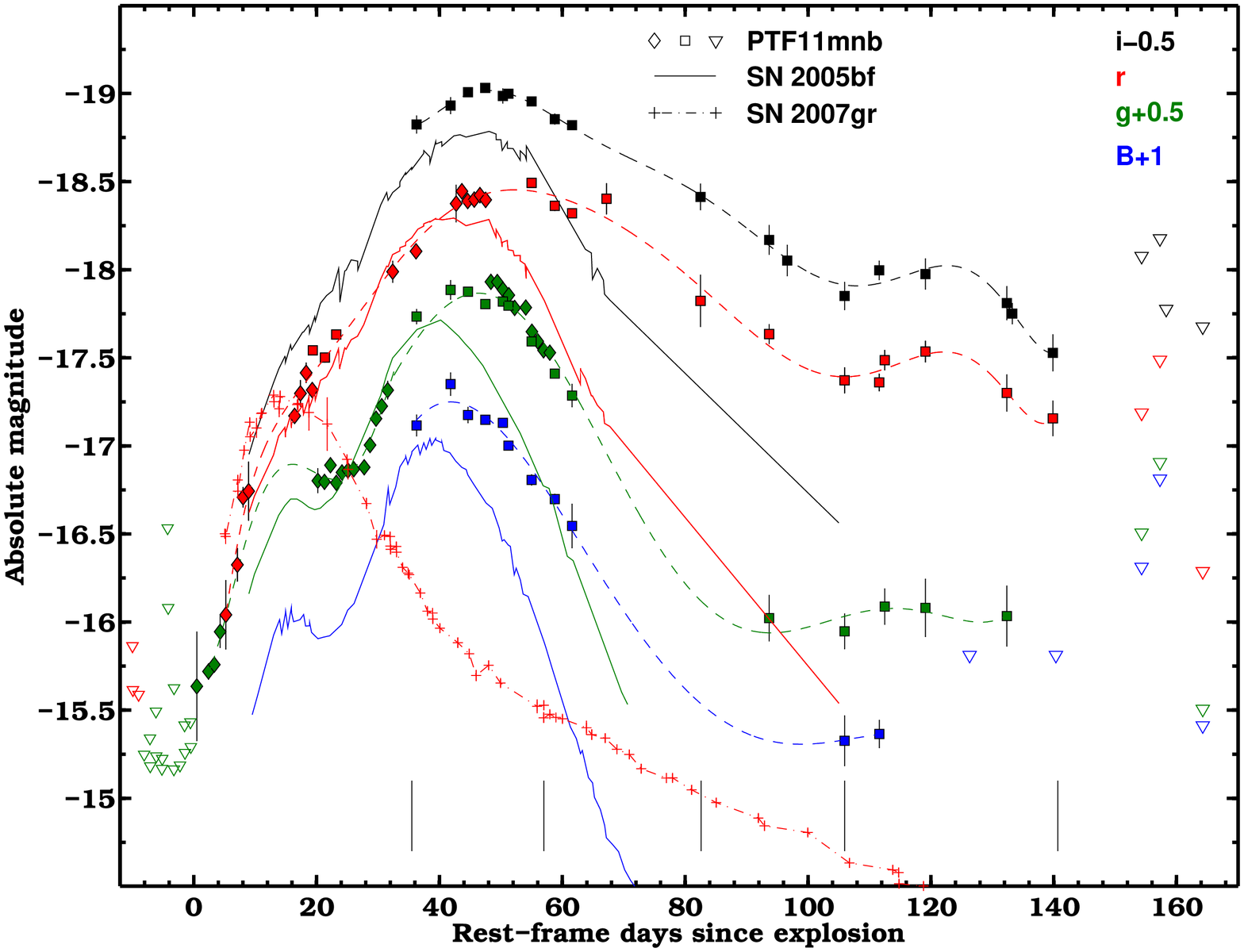}
\caption{\label{absmag} Optical ($Bgri$) light curves of
supernova PTF11mnb (colored symbols) compared to those of SN~2005bf
  \citep[solid lines,][]{folatelli06,bianco14}. The light curves of both events
  have been shifted 
in the same way, as indicated in the top-right corner. The five
spectral epochs for PTF11mnb are marked by vertical black segments at the bottom of
the plot. The light curves of PTF11mnb were fit with low-order
polynomials (dashed lines) to characterize their shape.
Limiting magnitudes are marked by empty triangles. P48 data are marked by diamonds, P60 data by squares. Until and including the
main peak, the light curves of SN~2005bf and PTF11mnb resemble 
each other, whereas after the main peak, PTF11mnb shows a much slower decline rate. We also compare the $r$-band light curves of PTF11mnb and SN~2005bf to the $R$-band light curve of the normal SN~Ic 2007gr \citep{hunter09}. While the first two SNe show the main peak at $\sim$52/42~d and at $-$18.5/$-$18.3 mag, SN~2007gr peaks at a similar phase (at $\sim$16~d) and magnitude ($\sim-$17.2 mag) as the early peak/bump of the other two events.}
\end{figure*}

\begin{figure}
\centering
\includegraphics[width=9cm]{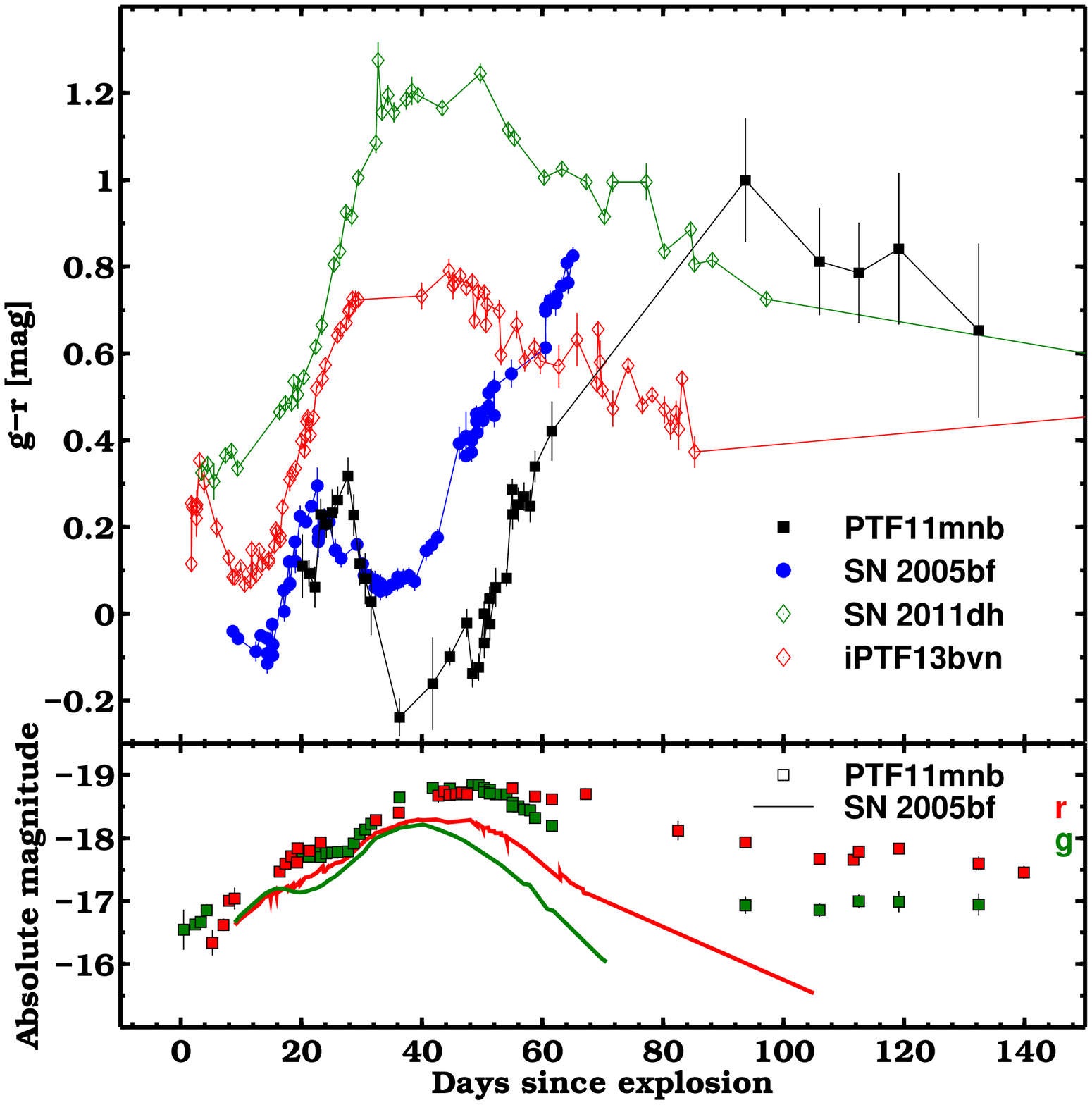}
\caption{\label{color}(Top panel) $g-r$ color comparison between PTF11mnb, SN~2005bf, SN~Ib iPTF13bvn \citep{fremling16} and SN~IIb 2011dh \citep{ergon14,ergon15}. The first two SNe reach a red peak in the color that coincides with the early peak in their light curves (see bottom panel). Then their $g-r$ becomes bluer reaching a minimum when the main peak of the light curves occurs. Thereafter, they rapidly evolve to redder colors. PTF11mnb also shows a shallow trend to the blue after $\sim$100~d, when the light curves are flatter (see bottom panel), and when SN~2005bf was not observed. iPTf13bvn  and SN~2011dh evolve very differently, reaching a $g-r$ minimum at peak magnitude, followed by a linear trend to the red and a final flatter color after $\sim$30~d. 
(Bottom panel) Absolute $g-$ and $r-$band light curves of PTF11mnb and SN~2005bf, plotted as a reference for the different phases of the color evolution. }
\end{figure}

In Fig.~\ref{absmag} we present the $Bgri$ light curves
of PTF11mnb. The first detection in $g$ band occured at $\sim-$15.63
mag. For 4 days (first three epochs) the light curve in the $g$ band shows an almost
constant value (within the errors). Thereafter, the SN was observed rising faster in $g$ band and imaged in
the $r$ band with P48. The $r$-band light curve shows a steep rise of $\sim$1.3 mag from $+$5.2~d to $+$17.4~d. This is followed by 
a flatter phase until 21.3 days. This phase is also seen in the $g$
band as an early peak/plateau between $\sim$20~d and 27~d. Following this early peak, both
$g$ and $r$ band rise to the main peak that occurs at $+$46.3~d and
$+$52.2~d, respectively. PTF11mnb was also observed in $B$ and $i$
band starting from $+$36~d. The $B-$ and $i-$band peaks occurred at
$+$41.9~d and $+$48.1~d, respectively. We determined the maxima by
fitting the light curves with low-order polynomials, 
marked in Fig.~\ref{absmag} by dashed lines.

The main peaks of PTF11mnb are quite broad (especially in the redder
bands), and are characterized by $\Delta m_{15}~=~$0.48, 0.53, 0.16, 0.24~mag in $B$, $g$, $r$, $i$, respectively.

After an almost linear decline that lasts until $+$106~d in $r$ and
$i$ band and $+$93~d in the $g$ band, the late light curves show a 
shallow rebrightening (0.1 mag in $r$ and $i$) that peaks at
$\sim$115~d in $g$ band and at $\sim$120~d in $r$ and $i$ band. The
last detection occurs after this shallow bump, at $+$140~d ($r$ and $i$) and
$+$132~d ($g$).

The comparison to the light curves of SN~2005bf reveals 
how these two events are similar, at least during the epochs 
up to and including the main peak. 
In Fig.~\ref{absmag} we show the 
absolute-magnitude light-curves of SN~2005bf from 
\citet{folatelli06} and \citet{bianco14} as solid colored lines. For SN~2005bf we 
assumed a distance modulus $\mu_{05bf}~=~$34.561~mag, 
redshift $z~=~$0.018913 (from NED), explosion epoch JD$=$2453458, 
galactic extinction $A_{B}(05bf)~=~$0.163~mag, $A_{g}(05bf)~=~
$0.148~mag, $A_{r}(05bf)~=~$0.102~mag, and $A_{i}(05bf)~=~
$0.076~mag 
(redshift from NED\footnote{NASA/IPAC Extragalactic Database: \href{https://ned.ipac.caltech.edu}{https://ned.ipac.caltech.edu}}, distance modulus computed assuming the same cosmology as used for PTF11mnb, extinction from \citealp{sf11}). Both the early $r-$band rise and the early $g-
$ and $r-$band peaks are 
similar in phase and absolute magnitude. However, the early peak in $r$ band is more pronounced in PTF11mnb than in SN~2005bf, where the $r$-band light curve shows a change in curvature rather than a clear maximum. In $g$ band, both SNe show clear early maxima. 
SN~2005bf shows 
its main peak in $B$, $g$ and $r$ band slightly earlier (3--10~d) than PTF11mnb, whereas the peak epoch is very similar in $i$ band. The $Bgri$ peaks of SN~2005bf are only 0.2--0.3 mag fainter than those of PTF11mnb. 

After the main peak, the light curves of SN~2005bf 
decline much faster than those of PTF11mnb. At 100 days, SN~2005bf has declined by $\sim$2.5 mag from peak in $r$ band, whereas PTF11mnb has declined by merely $\sim$1.0 mag in the same band.

We summarize the main light curve properties of PTF11mnb in Table~\ref{tab:propgal}.

The color evolution of PTF11mnb is similar to that of SN~2005bf, as
shown in Fig.~\ref{color}. The early light-curve peak (see bottom
panel) corresponds to an early maximum in the $g-r$ colors, whereas
the main light curve peak correspond to a minimum in the $g-r$
colors. These minima are followed by a rise in the $g-r$ values, in
both SNe. PTF11mnb is slightly redder than SN~2005bf around the early
peak, although it is bluer later on. 
For PTF11mnb we have late detections in $g$ and $r$, and at epochs later than 90 days $g-r$ slowly becomes bluer. 
The $g-r$ colors of normal SE~SNe, such as those of iPTF13bvn \citep{fremling16} and SN~2011dh \citep{ergon14,ergon15} shown in the figure, evolve differently. After the $g-r$ minimum in correspondence with the light-curve peak, the $g-r$ of these normal SE SNe becomes redder until $\sim$30~d and then slowly turns bluer. 

\section{Spectra}
\label{sec:spec}

\begin{figure*}
\centering
\includegraphics[width=18cm]{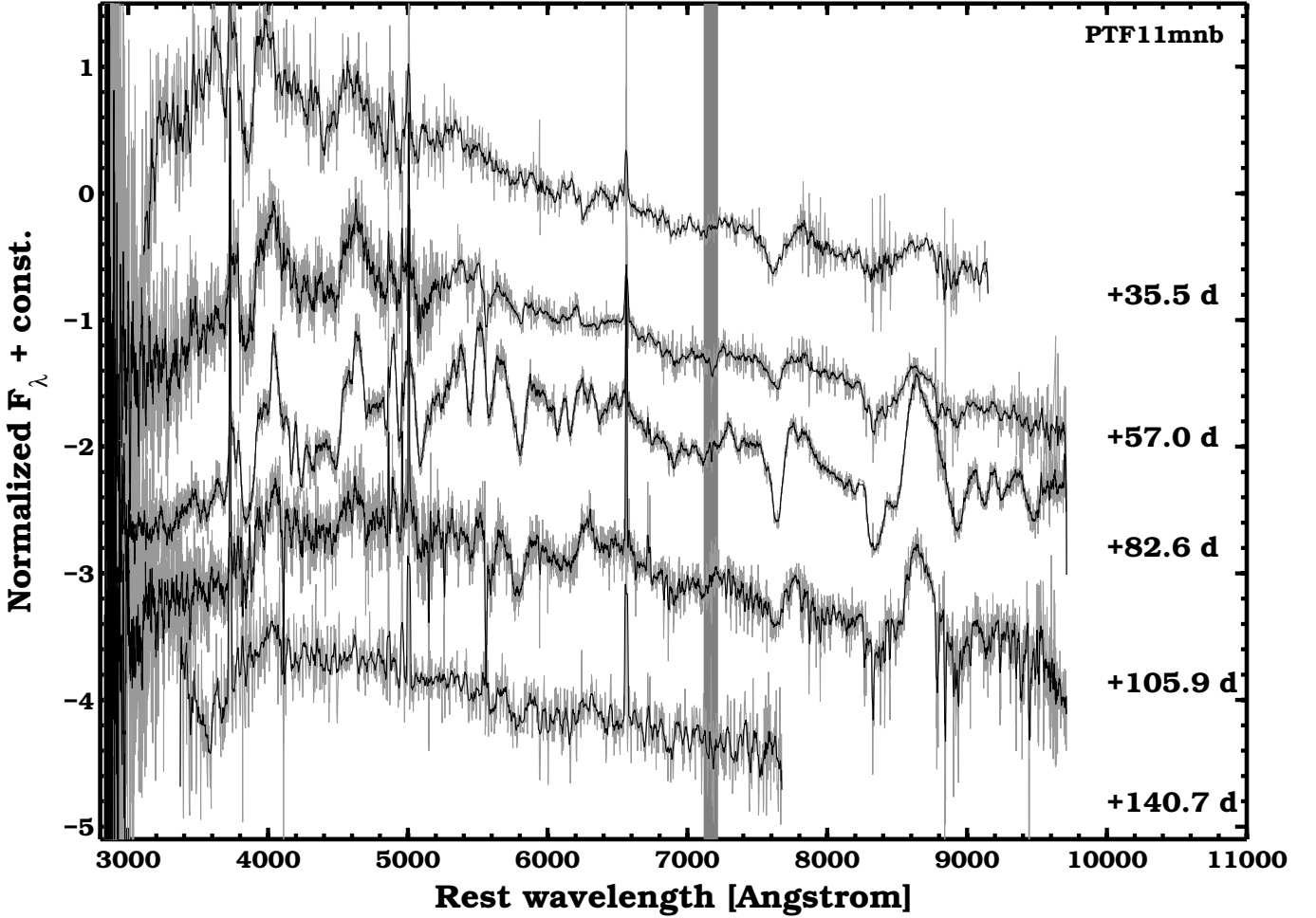}
\caption{\label{spec}Spectral sequence of PTF11mnb. Phases in rest-frame days since explosion are reported next to each spectrum. A gray area masks the strongest telluric feature at 7600~\AA.}
\end{figure*}

\begin{figure*}
\centering
\includegraphics[width=17.5cm,height=13cm]{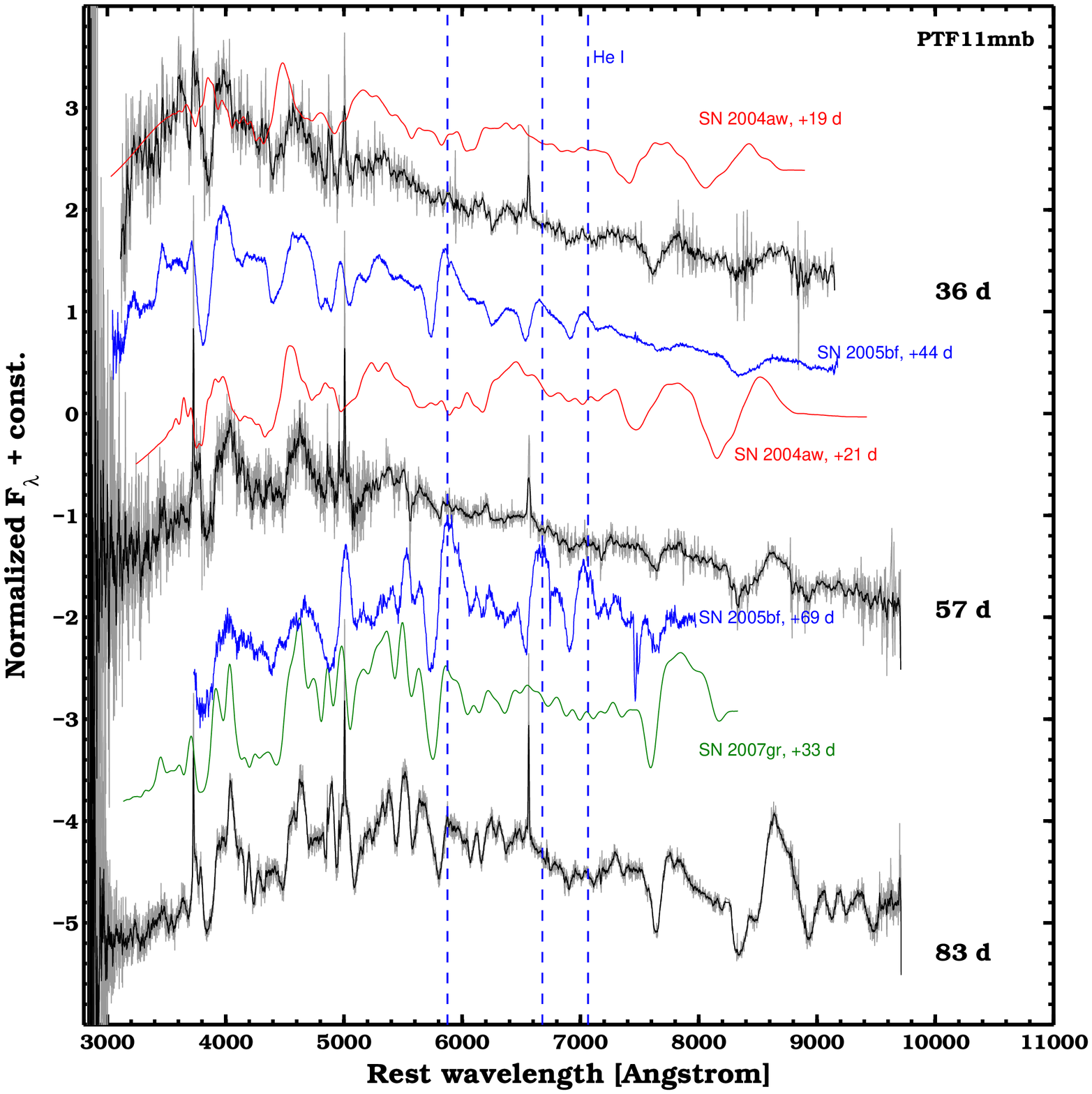}\\
\includegraphics[width=17.5cm,height=9cm]{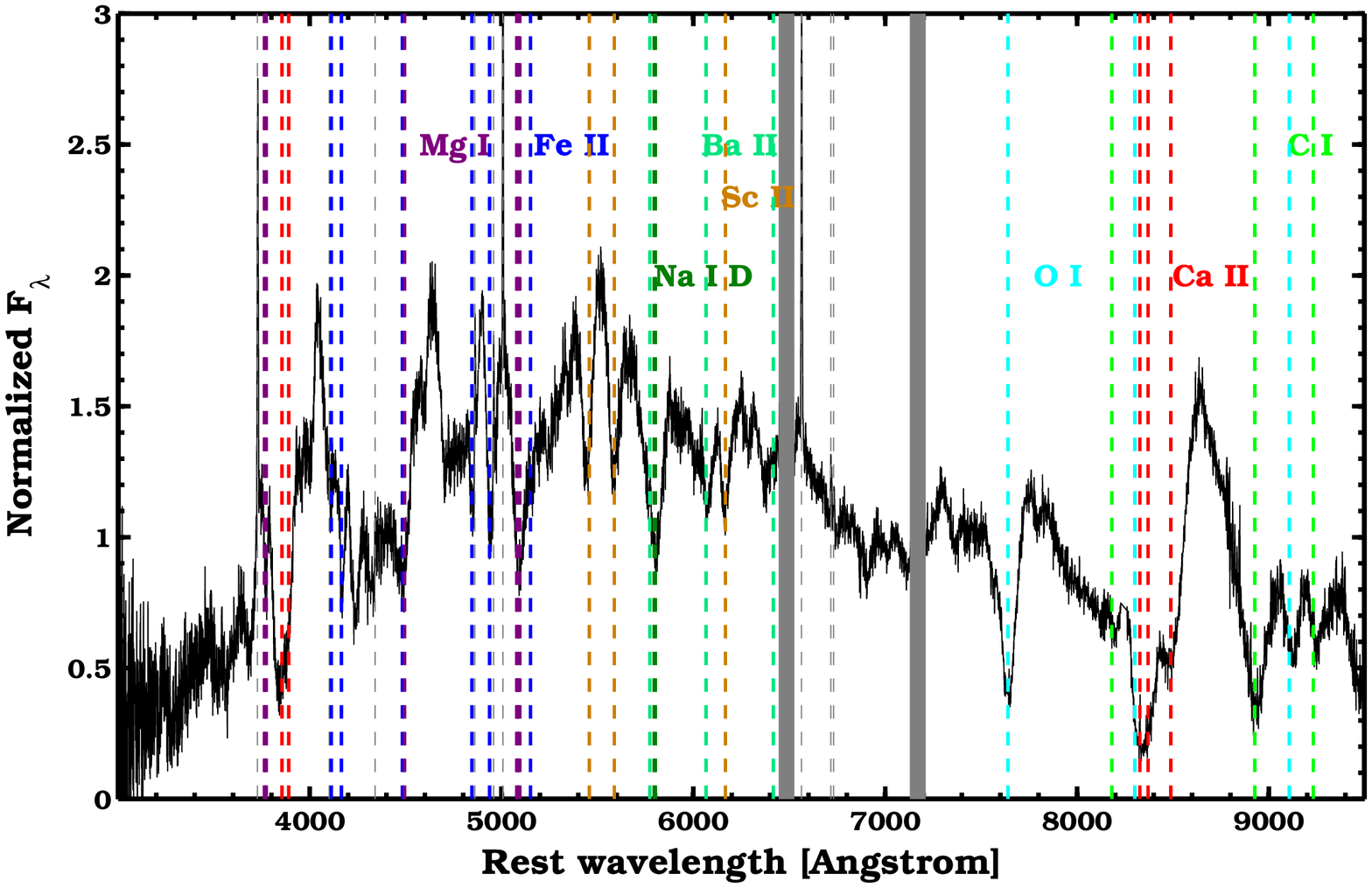}
\caption{\label{speccomp}(Top panel) Spectral comparison of PTF11mnb to other SNe~Ic (from SNID templates, \citealp{snid}; \citealp{liu14}) and SN~2005bf \citep{folatelli06}. Phases from explosion in rest-frame days are reported next to each spectrum. SN~2005bf is the only SN showing strong He lines, which are not present in PTF11mnb. (Bottom panel) Line identification on the best PTF11mnb spectrum. Typical SN~Ic features dominate the spectrum, with ions such as \ion{Ca}{ii}, \ion{O}{i}, \ion{Fe}{ii},  \ion{Mg}{i}, \ion{C}{i}, \ion{Na}{i}. Lines from ions such as \ion{Sc}{ii} and \ion{Ba}{ii} are also visible. Dashed-lines of the same color mark the same ion, which is labeled in the plot with the same color. Gray areas mark the most prominent telluric features.}
\end{figure*}

PTF11mnb was observed five times in spectroscopy, from $+$35~d until
$+$141~d. The spectral sequence is shown in Fig.~\ref{spec}. The spectra will be available via WISeREP \citep{yaron12}.

 We did not removed the narrow emission lines due to the host galaxy from the SN spectra. These are well visible in Fig.~\ref{spec}. We can observe narrow H$\alpha$, H$\beta$, [\ion{O}{iii}]~$\lambda\lambda$4959,5007, [\ion{O}{ii}]~$\lambda$3727, and \ion{S}{ii}~$\lambda\lambda$6717,6731 emission lines. Beside the narrow emission lines, the host-galaxy continuum emission (which peaks in the blue, see Fig.~\ref{fc}) could not be completely removed and it is likely contaminating the continuum of the spectra, affecting their shape in particular at late epochs.

The spectra of PTF11mnb are similar to other SN~Ic spectra, as illustrated in Fig.~\ref{speccomp} (top panel). Some of the best SN~Ic spectral fits obtained with SNID\footnote{Supernova Identification:\\ \href{https://people.lam.fr/blondin.stephane/software/snid/}{https://people.lam.fr/blondin.stephane/software/snid/}} \citep{snid} are shown for comparison, and they clearly match the features of PTF11mnb. 
On the other hand, the spectra of PTF11mnb are remarkably different from those of SN~2005bf (see Fig.~\ref{speccomp}, top panel), which show increasingly stronger \ion{He}{} lines. PTF11mnb never became a SN Ib.

The  spectra of PTF11mnb show the 
characteristic lines of a SN~Ic dominating over the continuum from the
second spectral epoch. We provide line identifications using the
spectrum with best signal (+83~d) 
shown in Fig.~\ref{speccomp} (bottom panel) and comparing with line identifications of SE~SN spectra from the literature \citep{taubenberger06,parrent07,hunter09}. 
Neither \ion{H}{} nor \ion{He}{} lines are detected. The spectrum is
dominated by \ion{Ca}{ii}, \ion{O}{i} and \ion{Fe}{ii} lines. \ion{Ca}{ii}
and \ion{O}{i} characterize the red part of the spectrum, with P-Cygni
profiles. Lines from \ion{Na}{i}~D, \ion{Mg}{ii}, and \ion{C}{i} are also visible, as observed in other SE~SNe \citep[see e.g.,][]{elmhamdi06}. We can also see lines due to \ion{Sc}{ii} and \ion{Ba}{ii}, which are not very common in SE~SNe, but appear in SNe~II \citep[e.g.,][]{taddia16long}.

\begin{figure}
\centering
\includegraphics[width=9cm]{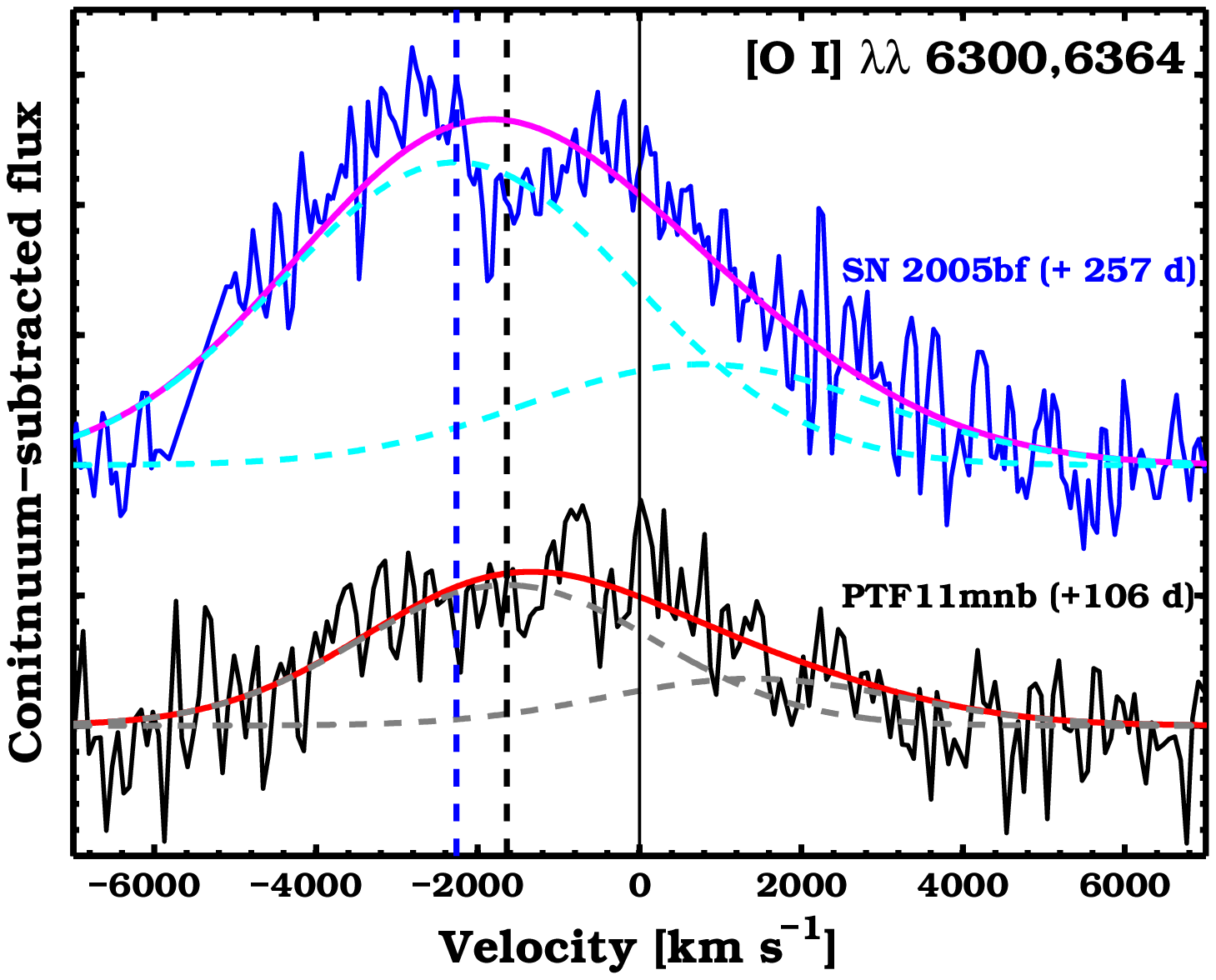}
\caption{\label{OI}[\ion{O}{i}]~$\lambda\lambda$6300,6364 lines from the fourth
  spectrum of PTF11mnb (+106 d, black line) as compared to the same lines in a late spectrum of SN~2005bf \citep[from][blue
  line]{modjaz08}. The [\ion{O}{i}]~$\lambda\lambda$6300,6364 lines are shown in velocity space and continuum subtracted, after fitting the continuum with a low order polynomial. The lines are fitted with the sum (solid curves) of two Gaussians (shown as dashed curved), with the same FWHM, relative flux ratio of 1:3, and peaks separated by 64~\AA. 
   The [\ion{O}{i}]~$\lambda$6300 line peaks are marked by thick dashed lines,  at
  $-$1640 km~s$^{-1}$ for PTF11mnb and $-$2260 km~s$^{-1}$ for
  SN~2005bf. The zero velocity of the line (at 6300~\AA) is marked by a solid,
  vertical black line.}
\end{figure}

In the last two spectra, [\ion{O}{i}]~$\lambda\lambda$6300,6364 is observed in
emission. In Fig.~\ref{OI} we observe that the emission lines [\ion{O}{i}]~$\lambda\lambda$6300,6364 detected in the spectra of PTF11mnb at $>100~$d are blueshifted. After continuum subtraction, the shape of these two lines can be reproduced by the sum of two Gaussians with the same FWHM, with peaks at 64~\AA\ from each other, and having relative flux ratio of 1:3 \citep[see][for a similar analysis]{mili10}. This is also the case for the  [\ion{O}{i}]~$\lambda\lambda$6300,6364 lines in the nebular spectrum of SN~2005bf from \citet{modjaz08}, which we show in Fig.~\ref{OI}. The individual Gaussian components are shown by dashed curves, the total best fit with solid curves for both SNe. 
The vertical dashed lines (black for PTF11mnb, blue for SN~2005bf) indicate that the blueshift is larger for SN~2005bf ($-$2260 km~s$^{-1}$) than in PTF11mnb ($-$1640 km~s$^{-1}$). We notice that, 
apart from the [\ion{O}{i}] emission lines, the spectra of PTF11mnb are not fully nebular even at these late epochs.

\begin{figure}
\centering
\includegraphics[width=9cm]{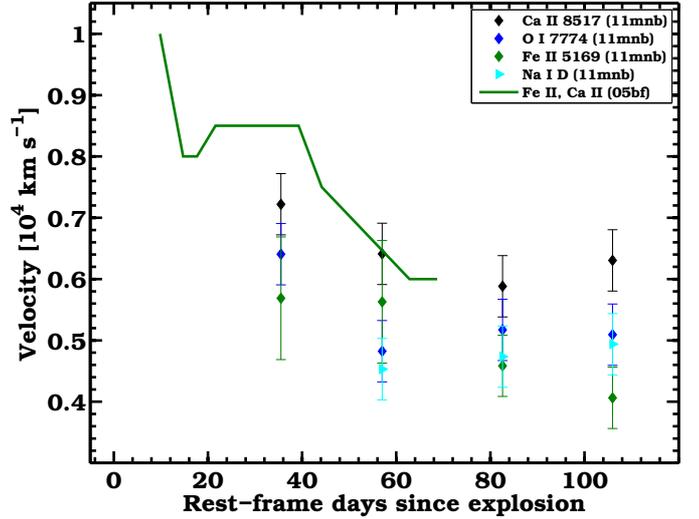}
\caption{\label{vel}Expansion velocities of PTF11mnb as measured from the absorption minima of some spectral lines showing P-Cygni profiles. We also report the \ion{Fe}{ii} and \ion{Ca}{ii} velocities of SN~2005bf for comparison. SN~2005bf also shows high and low velocity components for these lines, as shown by \citet{folatelli06}.}
\end{figure}

From the first four spectra we measured the P-Cygni absorption velocities of some
of the most prominent lines. We report them in
Fig.~\ref{vel}. \ion{Ca}{ii} shows a constant velocity of about
6000~km~s$^{-1}$, \ion{O}{i}~$\lambda$7774 is detected at about the
same velocity in the first spectrum and later on at
$\sim$5000~km~s$^{-1}$. A similar velocity is shown by \ion{Na}{i}~D
and \ion{Fe}{ii}~$\lambda$5169. The expansion velocities of PTF11mnb as measured from the \ion{Fe}{ii}~$\lambda$5169 line are slower than those of normal SNe~Ic \citep{modjaz16} at similar epochs ($\sim$4500~km~s$^{-1}$ instead of $\sim$7000~km~s$^{-1}$). SN~2005bf
display higher velocities than PTF11mnb when we
compare the \ion{Fe}{ii} lines, and PTF11mnb does not have spectra taken early enough to check if there were fast velocity components in \ion{Fe}{ii} and \ion{Ca}{ii} as observed in SN~2005bf (see \citealp{folatelli06}).

\section{Modeling}
\label{sec:model}

\subsection{Bolometric properties}

\begin{figure}
\centering
\includegraphics[width=9cm]{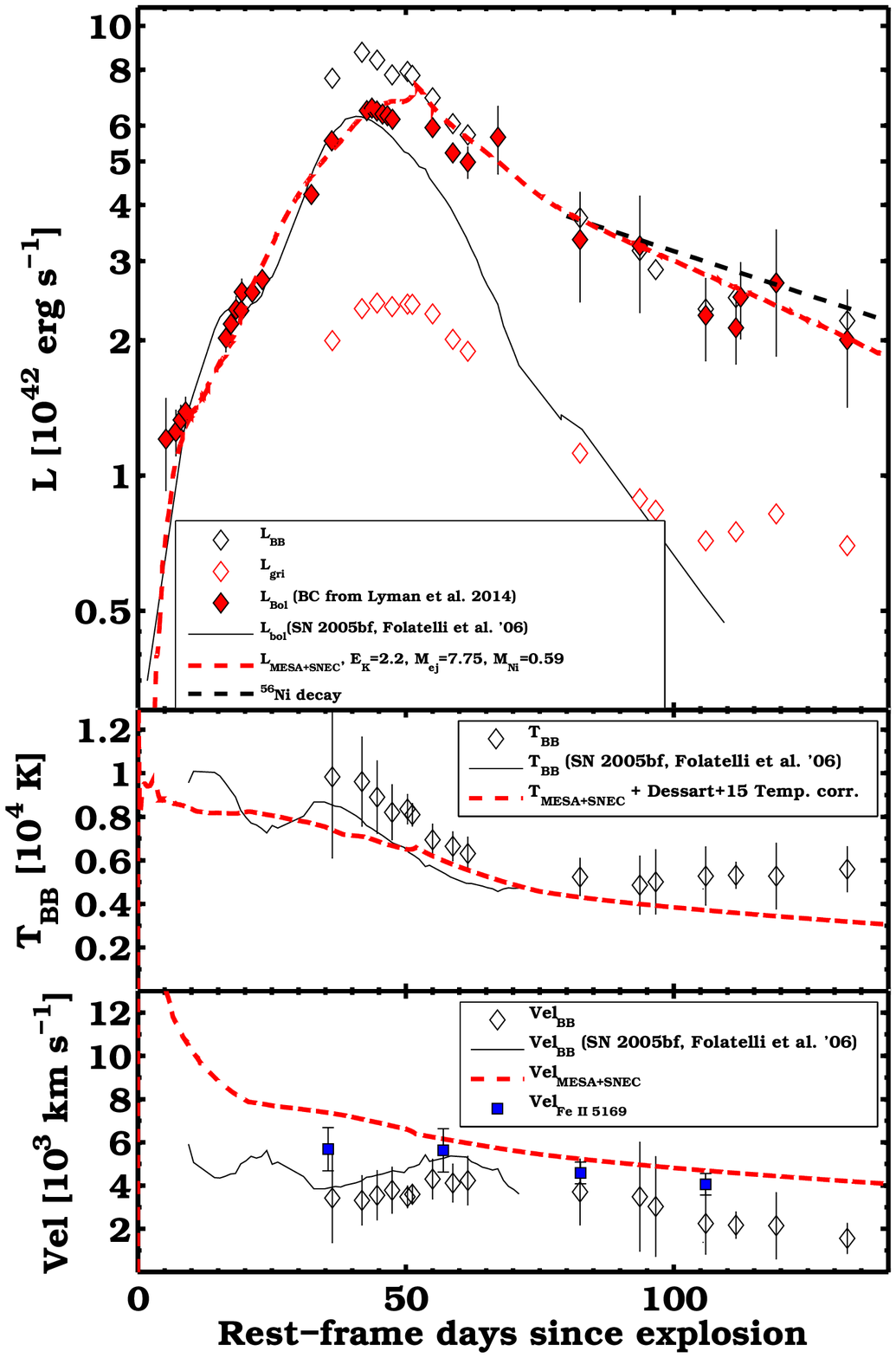}
\caption{\label{LTR}Top panel: Bolometric light curve of PTF11mnb. The luminosity from the integrated $gri$ SEDs, from the BB fit to the SED, and from $g$ and $r$ along with the bolometric corrections by \citet{lyman14} are marked by empty red diamons, empty black diamonds and red diamonds, respectively. The light curve of SN~2005bf from \citet{folatelli06} is plotted as a black solid line. The luminosity from the $^{56}$Co decay (M[$^{56}$Ni] $=$ 0.59 $M_{\odot}$) is marked by a black dashed line. The preferred  hydrodynamical model is shown as a red dashed line. Central panel: Black-body temperature evolution of PTF11mnb (black diamonds) as compared to that of SN~2005bf from \citet{folatelli06} and that of the preferred hydrodynamical model (red dashed line). Bottom panel: BB velocity evolution of PTF11mnb (black diamonds) as compared to that of SN~2005bf from \citet{folatelli06}, to the photospheric velocity of the preferred hydrodynamical model (red dashed line) and to the \ion{Fe}{ii} P-Cygni absorption velocity of PTF11mnb (blue squares).}
\end{figure}

In order to build a quasi-bolometric light curve of PTF11mnb, 
we make use of its broadband photometry. We interpolate the $gri$ light curves to the epochs of the P60 $r$ band. Then we convert the interpolated magnitudes (corrected for the extinction) into fluxes, at the effective
wavelength of the corresponding filters \citep{fukugita96}. The
resulting spectral energy distributions (SEDs) are integrated and
multiplied by 4$\pi$D$^2$, where $D$ is the luminosity distance to the
supernova. By doing this, we obtain $L_{gri}$, which is marked by empty red diamonds in Fig.~\ref{LTR} (top panel). To account for the 
emission at bluer and redder wavelengths, we fit the $gri$ SEDs with
a black-body (BB) function, which we then integrated to derive $L_{BB}$ (empty black diamonds in Fig.~\ref{LTR}, top panel). As the BB fit tends to overestimate the flux in the blue, where metal lines typically absorb a significant fraction of flux, we resort to use the bolometric corrections by \citet{lyman14} along with $g$ and $r$ band in order to derive the final bolometric light curve ($L_{Bol}$, red diamonds in the top panel of Fig.~\ref{LTR}). By using only $g$ and $r$ along with the bolometric corrections, we also produce the rising part of the bolometric light curve, since only those two filters were used at that phase. We notice that, among other events, also SN~2005bf was used by \citet{lyman14} to contruct the bolometric corrections for SE SNe that we use here for PTF11mnb. 
Our bolometric light curve is compared to that of SN~2005bf from
\citet{folatelli06} in the top panel of Fig.~\ref{LTR}. The two light
curves are almost identical until and including peak, thereafter SN~2005bf drops much faster, whereas PTF11mnb seems to follow the decay rate of $^{56}$Co (marked by a black dashed line), with the slope of $M_{Bol}$ at epochs later than 70~d being 0.011$\pm$0.03 mag~d$^{-1}$.

As a byproduct of the BB fit to the SEDs, we also obtain the temperature and the velocity evolution (derived from the BB radius divided by the time since explosion) of PTF11mnb, reported in the central and bottom panels of Fig.~\ref{LTR} (black diamonds). This evolution is rather similar to that of SN~2005bf \citep{folatelli06}, which is marked by a solid black line. A maximum in the temperature is reached around the epoch of the second light curve peak in both SNe, as already suggested by the color evolution (Fig.~\ref{color}). The velocity has a peak around $\sim$60~d in both SNe.

\subsection{Double-peaked $^{56}$Ni distribution scenario}
\label{sec:double}
The bolometric light curve of PTF11mnb, as well as its velocity and
temperature evolution, can be reproduced by a model similar to
that suggested for SN~2005bf by \citet{tominaga05} and
\citet{folatelli06}. This considers the explosion of a massive star
characterized by a double-peaked distribution of $^{56}$Ni. A
relatively small fraction of $^{56}$Ni  in the outer layers makes it
possible to power the early peak. A larger fraction of $^{56}$Ni
situated deeper in the ejecta allows us to reproduce the main peak. In the
case of SN~2005bf, the post main-peak decline was fast and required an
artificially reduced gamma-ray trapping in this scenario. The
late-time photometry of SN~2005bf could not be reproduced with a radioactive
model, so a magnetar was invoked by \citet{maeda07}. Contrarily to
SN~2005bf, we do not need to artificially reduce the gamma-ray opacity
by a factor of 10 to match the post main peak bolometric light
curve. The slow decline of PTF11mnb is instead fully consistent with 
the decay rate of  $^{56}$Co (see black dashed line in Fig.~\ref{LTR}). 

As the first step to model PTF11mnb, we produced a massive (final mass 9.5~$M_{\odot}$) pre-SN star with the Modules for Experiments in Stellar Astrophysics (MESA; \citealp{paxton11}), similar to what was used by \citet{folatelli06} to model SN~2005bf. 
We started from a star with initial mass $M_{\rm ZAMS}=$85~$M_{\odot}$ and a slightly sub-solar metallicity (Z$=$0.01), consistent with what we inferred for the location of PTF11mnb (see Sect.~\ref{sec:hg}). We set the star rotation velocity to 350 km~s$^{-1}$. The rotation and the initial mass were adjusted to reproduce the desired final mass, and at the same time to strip the entire hydrogen envelope and the helium envelope. The final progenitor star model contains merely 0.6~$M_{\odot}$ of He in the outer part of the ejecta, which may be consistent with a SN~Ic progenitor (see for example the progenitor star model 5p11 in \citealp{dessart15}).

\begin{figure}
\centering
\includegraphics[width=9cm]{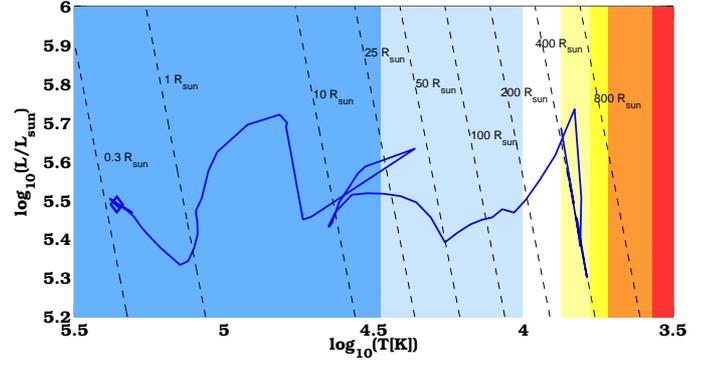}
\caption{\label{HR}HR-diagram evolution of the star produced with MESA and used to model the bolometric light curve of PTF11mnb with SNEC.}
\end{figure}

In Fig.~\ref{HR} we report the evolution of this star in the
Hertzprung-Russell (HR) diagram, until collapse. In the final stage of
its life this star sits in the Wolf-Rayet (WR) part of the HR diagram
(see blue diamond), with high luminosity  (10$^{5.5}$ solar
luminosity), high temperature (almost 10$^{5.36}$ K) and a compact radius (0.35~$R_{\odot}$).

\begin{figure}
\centering
\includegraphics[width=9cm]{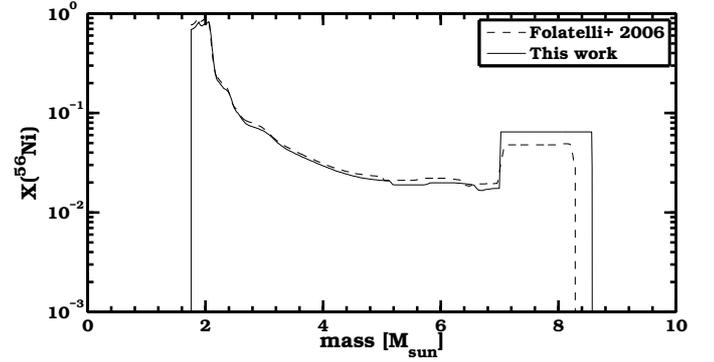}
\caption{\label{Ni_distrib}$^{56}$Ni distribution used for the best hydrodynamical model of PTF11mnb, as compared to that used for SN~2005bf by \citet{folatelli06}. Both are double-peaked, in order to reproduce  both the early and the main peak in the bolometric light curve.}
\end{figure}

 We exploded this star with the SuperNova Explosion Code (SNEC;
 \citealp{morozova15}). For an explosion energy of
 E~$=$~2.2$\times$10$^{51}$~erg (and a remnant mass of 
1.7~$M_{\odot}$,  
i.e. an ejecta mass of 7.8~$M_{\odot}$), and using the double-peaked $^{56}$Ni distribution plotted in Fig.~\ref{Ni_distrib} (which was slightly modified from the one of \citealt{folatelli06}), we reproduce the bolometric light curve of PTF11mnb rather accurately (see the thick red dashed lines in Fig.~\ref{LTR}). 
The total $^{56}$Ni mass of this model is 0.59~$M_{\odot}$. It is necessary to use a double-peaked $^{56}$Ni  distribution to  reproduce both the early and the main peak. We modified the SNEC code in order to include the non-uniform $^{56}$Ni distribution shown in Fig.~\ref{Ni_distrib}.

 The model is also able to  somewhat reproduce the photospheric-velocity profile as measured from the \ion{Fe}{II}~$\lambda5169$ line in the spectrum, as well as the temperature evolution. We notice that in order to directly compare the photospheric temperature from the SNEC model and the BB temperature from the fit of the $gri$ SEDs, we corrected the SNEC temperature profile by the ratio between the photospheric temperature and the $VI$ color temperature for SNe~Ibc presented by \citet{dessart15} in their table A3. 

An important assumption in the
hydrodynamical model is that related to the opacity floor. 
Following \citet{morozova15}, we have adopted a linear scaling of the opacity floor with metallicity. 
In particular, we started our model investigation with 0.01~cm$^{2}$~g$^{-1}$ at $Z=0.02$ (in the envelope) 
and 0.025~cm$^{2}$~g$^{-1}$ in the core, where $Z=1$. The opacity floor value in the envelope is the same as adopted 
for SNe~II and SNe~IIb in  \citet{bersten11} and \citet{ergon15}, respectively. The opacity floor in the 
core is the same as adopted for hydrogen-poor, and helium-rich SNe (see \citealp{ergon15}). This value was 
calibrated using a sophisticated radiative transfer code, STELLA \citep{blinnikov93}. However, we notice that the optimal
 value of the opacity floor for SNe~Ic, i.e., for helium-poor SNe, has not been presented in the literature yet. 
Therefore, we started our model investigation by assuming that the opacity floor in the core is similar for SNe~IIb and SNe~Ic, which are
both hydrogen poor. 
Our model with core opacity floor set to 0.025~cm$^{2}$~g$^{-1}$ succeeds in reproducing the velocity profile 
and the light curve.  However, around peak it shows a narrow feature 
in the light curve 
that is not observed in our SN. By slightly modifying 
the opacity floor in the core, i.e., adopting 0.04~cm$^{2}$~g$^{-1}$, we 
could smooth the narrow feature around 
the peak of the light curve model while preserving the good match to the velocity  profile from the \ion{Fe}{ii}~$\lambda5169$ line. For larger 
values of core opacity floor, the light curve match is still good until 0.07~cm$^{2}$~g$^{-1}$ and 
acceptable adopting 0.1~cm$^{2}$~g$^{-1}$. However, the velocities for the models with higher core 
opacity-floor values are above those observed in our SN. For even higher opacity-floor values, in particular  
if we use the core opacity floor for hydrogen-rich SNe~II presented by \citet{bersten11} (0.24~cm$^{2}$~g$^{-1}$) and 
used by \citet{morozova15}, the light-curve model instead shows an extremely long rise time, which is not compatible with the observed light curve (the other parameters being the same).

\subsection{Magnetar scenario}

\begin{figure}
\centering
\includegraphics[width=9cm]{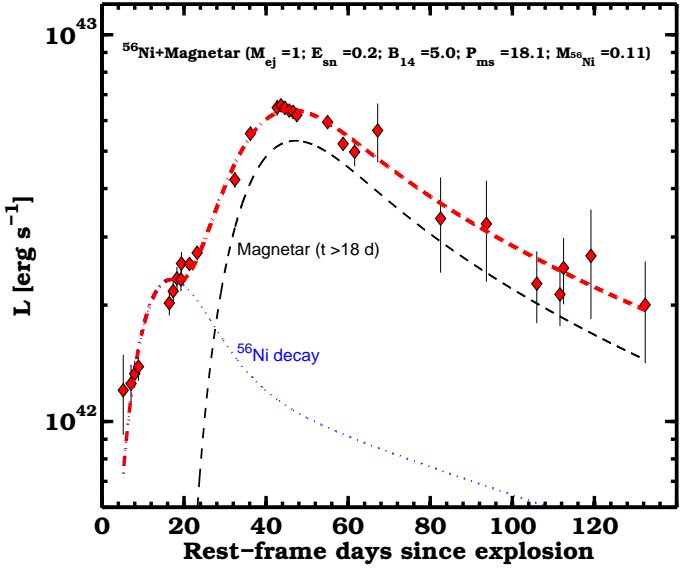}
\caption{\label{hybrid} Hybrid bolometric light curve model (red
  dashed line) of PTF11mnb. The first peak is fit by the radioactive
  decay of $^{56}$Ni and represented by an Arnett model (blue dotted
  line). The main peak is powered by a magnetar model (dashed black line) from \citet{kasen10}, which kicks in 18 days after exploson. The best fit parameters are reported at the top of the plot.}
\end{figure}

For SN~2005bf, \citet{maeda07} proposed a spinning magnetar as the main powering source of the light curve, based on the fast decline after the main peak, and on the late-time photometry. These properties were difficult to reconcile with a scenario where the radioactive decay was the powering mechanism.

If we fit the bolometric light curve after $+$30~d with a magnetar
model \citep{kasen10}, we obtain ejecta mass $M_{ej}=3~M_{\odot}$,
kinetic energy $E_{K}=0.6\times10^{51}$~erg, magnetic flux density
$B=4.3\times10^{14}~G$, rotation period $P=14.2$~ms. Here we have assumed an
expansion velocity at peak of 5600 km~s$^{-1}$ from the \ion{Fe}{ii} velocity,
and opacity $\kappa~=~0.2~{\rm cm^2~g^{-1}}$. This simple model
cannot reproduce the early peak, and therefore we fit only the later
epochs (however, including all the epochs provides almost identical
parameters). To reproduce also the early peak in the magnetar scenario, we
consider the early light curve as possibly produced by a magnetar-induced
shock breakout (SBO), as described by \citet{kasen16}.
Using the above derived parameters from the fit of the main peak with
the magnetar model, we can derive the time of the SBO peak and its
luminosity \cite[their eqs.~26 and 27]{kasen16}. Both these values are
off from what we actually observe in
PTF11mnb (we obtained a peak epoch for the SBO of 83 days, with luminosity 3.5$\times$10$^{43}$~erg), and therefore we do not favor a magnetar as the energy
source for the early peak. 

Since a magnetar-induced SBO model does not fit the early light curve
of PTF11mnb, whereas the early peak was nicely fit by a $^{56}$Ni
decay model (see Sect.~\ref{sec:double}), here we also consider a
hybrid model where the first peak is due to a regular SN explosion
powered by radioactivity, while the main peak is powered by a
magnetar. It is indeed possible to fit the bolometric light curve with
a simple Arnett model \citep{arnett82} representing the radioactive component, plus a
magnetar model \citep{kasen10} that injects energy in the ejecta
starting a few days after explosion. This is illustrated in
Fig.~\ref{hybrid}, where the Arnett model is plotted as a dotted blue
line and the magnetar model by a dashed black line. The sum of the two
models, shown as a dashed red line, fit well to the bolometric light
curve. The magnetar injection of energy would have to start after 18 days since the
SN explosion, otherwise it is not possible to reproduce both peaks
simultaneously.

 In such a scenario,
the SN explosion is characterized by $M_{ej}=1~M_{\odot}$, $E_{K}=0.2\times10^{51}$~erg,  $M_{^{56}Ni}=0.11~M_{\odot}$, and the magnetar by 
 $B=5.0\times10^{14}~G$ and $P=18.1$~ms. 
 We notice that the magnetar parameters in the
hybrid model are different from those of the magnetar-only model previously described. 
 Despite the nice fit of this hybrid model to the bolometric light curve, we still favor a progenitor scenario with the explosion of a massive star and a double-peaked $^{56}$Ni distribution, as it requires a single source of energy and less parameters, and particularly because it naturally fits the post main peak light curve. 

\subsection{Early light-curve modelling}
\label{sec:early}

\begin{figure}
\centering
\includegraphics[width=9cm]{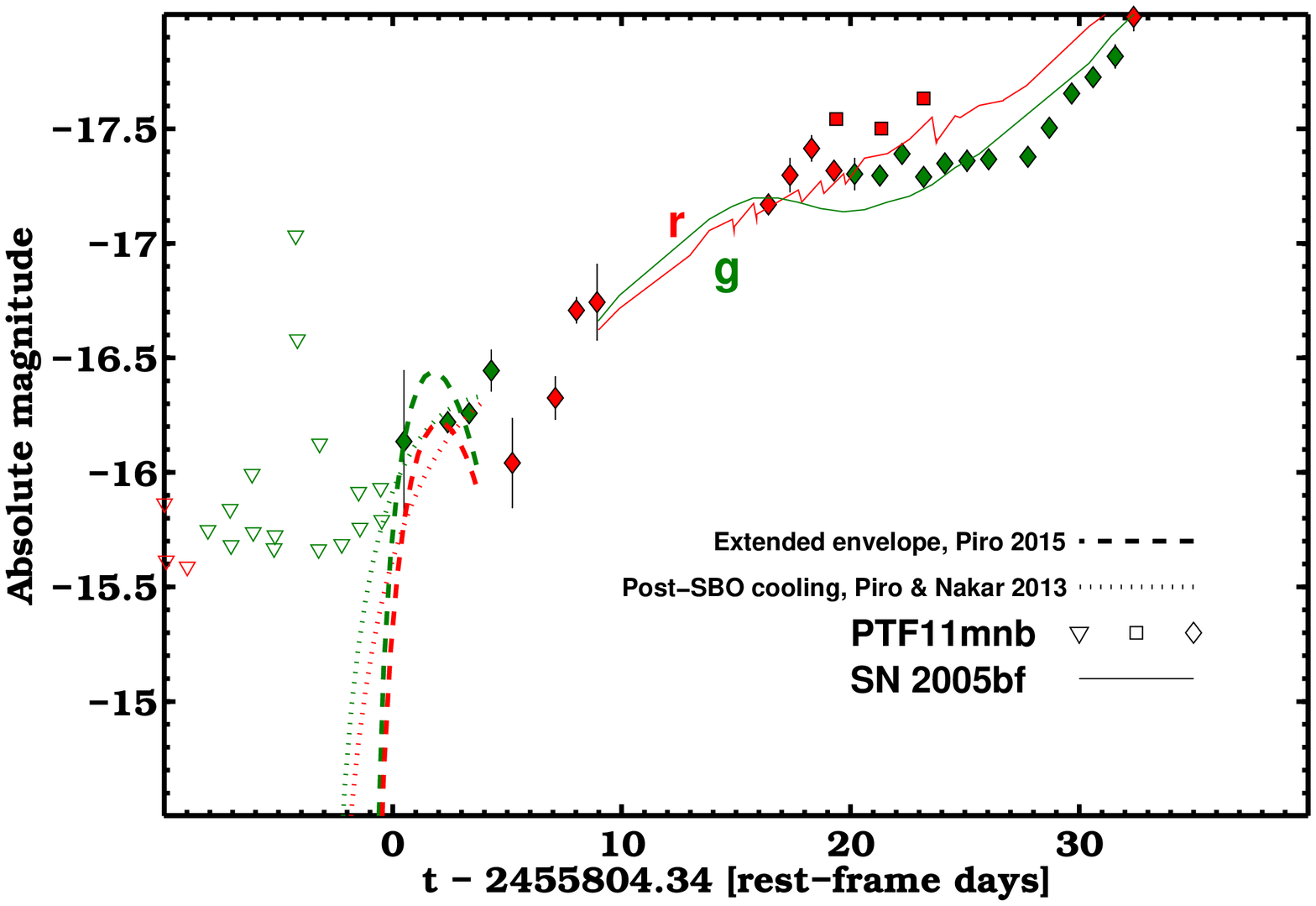}\\
$\begin{array}{cc}
\includegraphics[width=4.5cm]{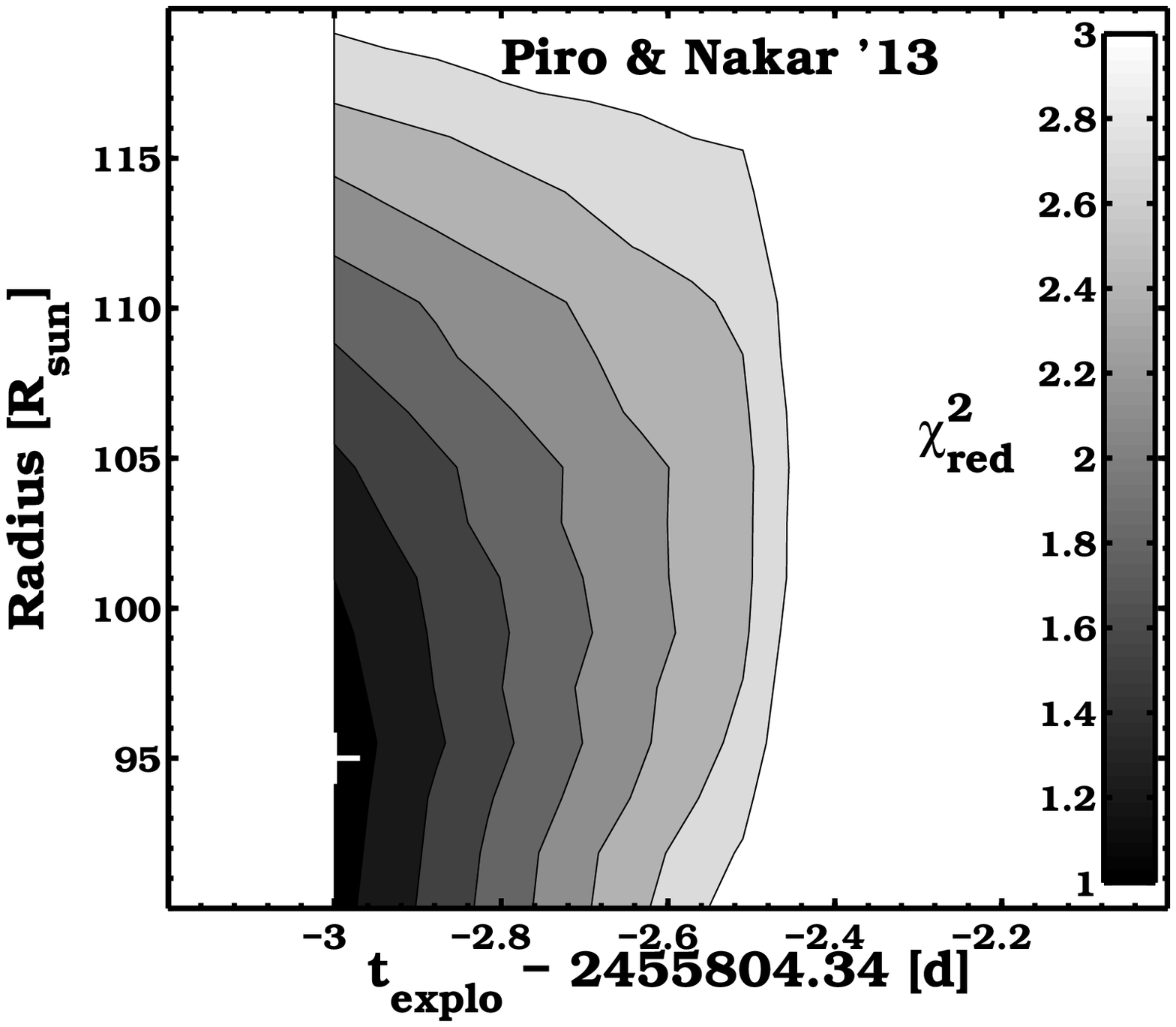}&
\includegraphics[width=4.5cm]{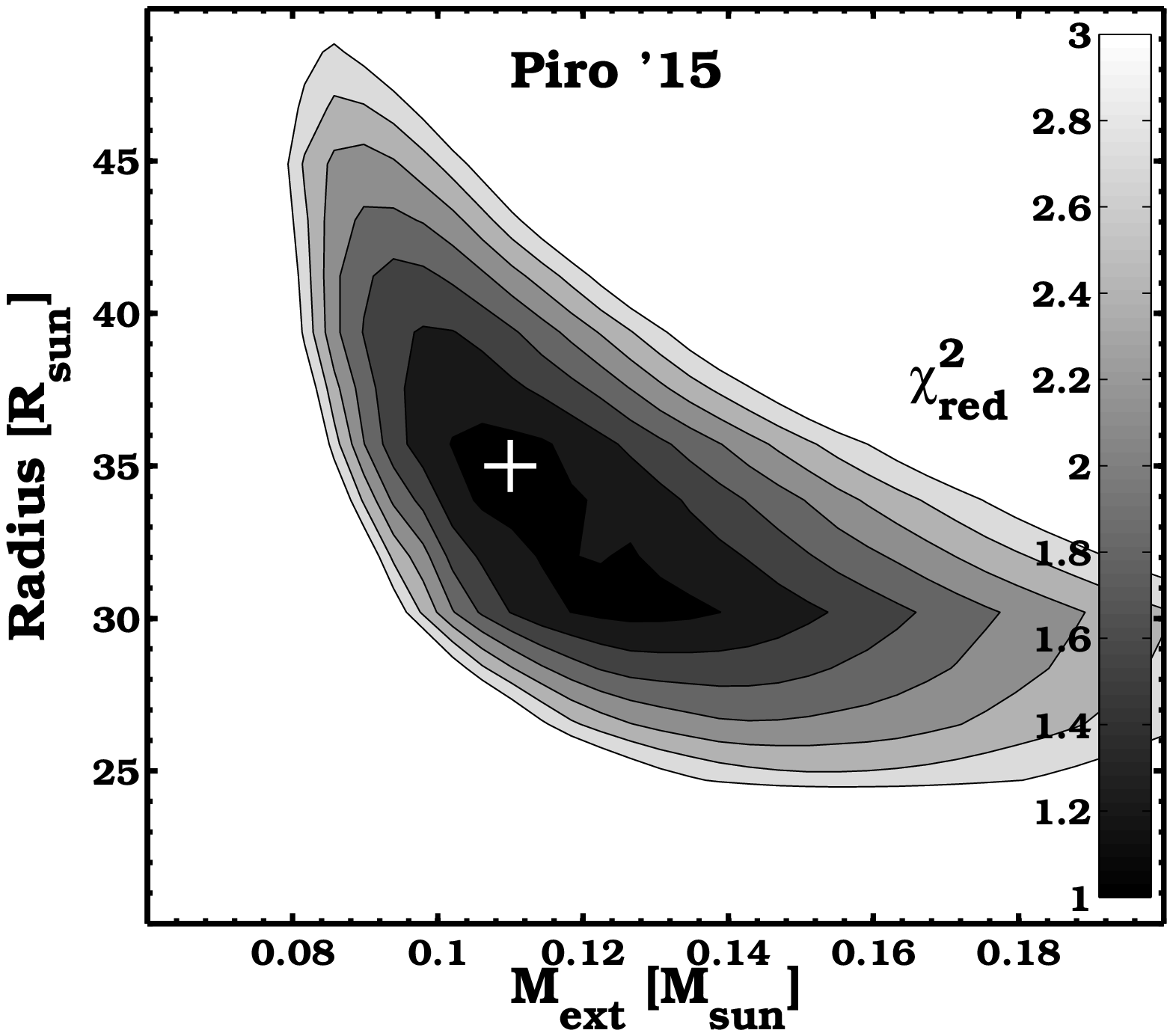}\\
\end{array}$
\caption{\label{fitearly}{\it Top panel:} The early $g$- and $r$-band
  light curves of PTF11mnb as compared to those of SN~2005bf.
Squares and diamonds mark P60 and P48 data, respectively. Triangles indicate upper limits.  
   PTF11mnb
  was discovered earlier, and it does not show a regular rising light
  curve in the first 3 epochs of $g$-band data. We fit these epochs
  with a shock breakout (SBO) cooling tail model from \citet{piro13}
  and with an extended envelope model from \citet{piro15}. The best fit of the latter model reveals the likely presence of a relatively
  extended envelope around the progenitor star
  ($\sim$35~$R_{\odot}$). {\it Bottom panels:} Reduced chi square for
  the two $g-$band light curve fits (on the left \citealp{piro13}, on the right \citealp{piro15}). 
When fitting the model by \citet{piro13} we solved for the explosion epoch and for the progenitor radius. When fitting the model provided by \citet{piro15} we solved for the extended envelope radius and mass (the explosion epoch was fixed to $-$1 d since the previously assumed value). The parameters are not highly degenerate and are well constrained by the early data. However, we lack multi-band coverage that would help to discriminate between the models.}
\end{figure}

The early $g-$band light curve of PTF11mnb appears rather flat in the
first three epochs, which occur merely $\lesssim$4 d after the last
non-detection. To reproduce this early emission, we adopt the model by
\citet{piro13} for the shock-break out cooling tail, assuming
$\kappa=0.2~\rm  cm^{2}~g^{-1}$, as well as kinetic energy and ejecta
mass from the hydrodynamical fit to the bolometric light curve. We
solve for the explosion epoch and for the progenitor radius, obtaining
a radius of $\sim$90--105 $R_{\odot}$ and an explosion epoch at
$\sim-$3~d since the average between discovery and last non-detection. This explosion epochs is consistent with the depth of the pre-discovery limits, as shown in Fig.~\ref{fitearly}. Explosions epochs earlier than that would not be compatible with the pre-detection limits.

The best fit for the first three epochs of $g$ band is shown in the top
panel of Fig.~\ref{fitearly} (green-dotted line). The corresponding
$r-$band light curve model (red-dotted line) is also shown, however the first epoch of $r$ band occurs too late to better constrain the radius and the explosion epoch. In the left-bottom panel of Fig.~\ref{fitearly} we show the reduced chi-square as a function of the progenitor radius and of the explosion epoch, from which we estimated the range of the best fit parameters. The best fit is shown by a white cross, and corresponds to a progenitor radius of 95~$R_{\odot}$. This is rather large compared to that of a regular WR star, which is about two orders of magnitude more compact. The progenitor star might have exploded while its radius was more extended than that of a WR star. This is also why the hydrodynamical model could not reproduce this early emission, as we exploded a compact (0.35~$R_{\odot}$) star.

However, another possibility is that an extended envelope of low mass surrounded the WR progenitor star at the moment of the explosion. This is how we interpreted the early emission from the explosion of the massive SN Ic progenitor of iPTF15dtg \citep{taddia16}. If we fit the model by \citet{piro15} that describes
the early emission from a star with an extended envelope (again
assuming the same opacity, ejecta mass, and explosion energy as in the previous model), we can solve for the radius and the mass of the outer envelope (and for the explosion epoch). We obtained an extended envelope of mass $\sim$0.11~$M_{\odot}$ and of radius $\sim$35~$R_{\odot}$. These parameters are rather well constrained by the fit (see Fig~\ref{fitearly}, right-hand bottom panel). Here the best explosion epoch was fixed to occur merely one day before the average between discovery and last non detection, to optimize the fit to the first $g$-band point. The best fit of the extended-envelope model to the $g-$band light curve is shown in Fig.~\ref{fitearly} as a dashed green line. The corresponding $r$-band model is shown with a red dashed line and it is not in contradiction with the first $r$-band observation, which however occurs to late to be used for further constraints.  

Discriminating between the two models is not easy as we lack multi-band coverage at that phase. With additional $r$-band observations, we could have constrained the temperature and observed if $g-r$ was more similar to what is predicted by the extended-envelope model \citep{piro15}, or by the shock-breakout cooling model \citep{piro13}. We could do that for iPTF15dtg \citep{taddia16}, where we had multi-band coverage. We favor the extended-envelope model as the fit is slightly better and because PTF11mnb is somewhat similar to iPTF15dtg and SN~2011bm, where we inferred the presence of an extended envelope.

There could also be the possibility that the early emission
is due to the afterglow of a long-duration GRB. We performed a search
for a possible GRB association via the interplanetary network (IPN),
as in \citet{taddia16}. In a longer interval that goes from seven days before discovery to
discovery, seven sources are found in the catalog. However, when we cross-checked with the INTEGRAL archive for the source position, none of these sources turned out to be compatible with PTF11mnb. 

\section{Discussion}
\label{sec:discussion}

The post main-peak photometry of SN~2005bf seems to 
favor a magnetar as the powering source for that event \citep{maeda07}. 
In the case of PTF11mnb, its slow decline rate instead  trace the $^{56}$Co decay, with no need for adjusting the gamma-ray escape fraction (as in SN~2005bf, see e.g., \citealp{tominaga05}). 

A magnetar cannot be excluded as the main source of the second peak, but $^{56}$Ni decay appears to be a simpler explanation. 
In favor of the magnetar scenario we might consider the shallow bump observed in the redder bands at $\sim$120~d, since 
a rapidly rotating central engine might produce some variability in
the light curve (see e.g., \citealp{nicholl16}). 
However, a shallow bump in the redder light curves after peak could be produced by the transition of \ion{Fe}{iii} ions to \ion{Fe}{ii} ions as in SNe Ia \citep[see also][]{roy16}. We do not have any signs of circumstellar interaction at late epochs that might explain this bump. 

As outlined in the study presented in \citet{taddia16}, with our
hydrodynamical model we did not intend to explore the entire parameter
space, but rather fixed a few parameters based on the previous models of SN~2005bf and on the information about PTF11mnb (in particular the ejecta mass, the helium-poor composition, the metallicity of the SN location, and the double-peaked $^{56}$Ni distribution). 

No asymmetry was taken into account when modeling the bolometric light
curve. The nebular oxygen line at 6300 \AA\ does suggest a slightly lower degree of asymmetry as compared to the case of SN~2005bf, however the double-peaked $^{56}$Ni distribution is likely produced by jets (e.g., see \citealp{tominaga05} about SN~2005bf). 

With PTF11mnb we found indications that some SNe~Ic might come from massive ($M_{ZAMS}~>~85~M_{\odot}$) progenitors characterized by an early emission compatible with the presence of extended material surrounding the progenitor star. The presence of material surrounding the SN at the moment of explosion might be explained in different ways. There are models pointing to massive WR stars  that can inflate a tiny quantity of mass thereby producing a halo structure around the progenitor core (e.g., \citealp{ishii99}). However, in these cases the inflated mass is order of magnitudes lower than what we inferred for PTF11mnb. 

Another explanation could be that this material is the result of a strong progenitor eruption. Enhanced mass loss due to eruptions just prior to explosion has been inferred for several SN progenitors (e.g., \citealp{ofek14}) and there are also models predicting these events \citep{shiode14}. In the cases of the SE~SNe 2006jc \citep{pastorello07} and PTF11qcj 	\citep{corsi14} eruptions were observed before the final explosion. If we assume that a wind with a velocity of 10$^3$ km~s$^{-1}$ from the progenitor produced the shell (a value that is typical for WR stars winds), then an extremely high and nonphysical (above 10$^2$ $M_{\odot}$~yr$^{-1}$) mass-loss rate lasting about 7 hrs would be needed to allocate $\sim$0.11~$M_{\odot}$ at a distance of $\sim$35~$R_{\odot}$. However, in the case of an eruption, the envelope might not be in hydrostatic equilibrium with the star (see \citealp{nakar14}, and \citealp{piro15}), therefore in this case we cannot asses its mass with precision, and therefore the estimate of the mass-loss rate would be unreliable. 
Also, if we assume lower eruption velocities the mass-loss rate would be more physical (e.g.,  $\sim$1~$M_{\odot}$~yr$^{-1}$ for about 1 month if we assume a velocity of 10 km~s$^{-1}$ ). We note that,  according to the models of \citet{shiode14}, 10$^{-3}$--1~$M_{\odot}$ can be inflated up to 10--100~$R_{\odot}$ in a time scale of month to decades, which would be in agreement with the last estimate. 

The envelope might also be produced during a common envelope  phase in a binary scenario \citep{chevalier12}, as it cannot be excluded that our massive progenitor was part of a binary system.

PTF11mnb is the only 2005bf-like SN in the PTF/iPTF sample of CC SNe. One single case corresponds to an observed fraction
of $\sim$0.12\% of the CC SNe and 0.51\% of the SE SNe (including
SNe~IIb) within (i)PTF. Given their peak luminosity and relatively
long rise time, 2005bf-like events should be easier to discover than 
normal SE SNe. This means they are likely even rarer. 

If we assume that only stars with $M_{ZAMS}$ $\gtrsim$85~$M_{\odot}$
can give rise to these events, and adopt a normal Salpeter initial
mass function (IMF), then we obtain in the single star progenitor
scenario that $\sim$4\% of CC SNe are formed from stars above 
that initial mass. Therefore, to match the rate of these events, 
some additional special conditions must occur in $\sim$3\% of these massive stars for a SN~2005bf like event to occur. It could also be that most of these very massive stars produce failed SNe, as concluded by \citet{smartt15}.

Ultimately, the similarities between PTF11mnb and SN~2005bf might indicate that the two events share a common powering mechanism, which the analysis of PTF11mnb suggests to be a double-peaked $^{56}$Ni distribution. SN~2005bf would be different compared to PTF11mnb after peak due to a larger gamma-ray escape, possibly related to the geometry of the explosion.
However, it is not excluded that both SN~2005bf and PTF11mnb are powered by a magnetar as proposed by \citet{maeda07} for SN~2005bf, with the magnetar fortuitously mimicking the $^{56}$Ni decay in the case of PTF11mnb. 

\section{Conclusions}

We have presented observations for PTF11mnb, a SN Ic whose early and
main-peak light curves resemble those of SN~2005bf. 
PTF11mnb never shows \ion{He}{} in its spectra, contrarily to SN~2005bf.

Its slowly declining light curve suggests that a doubled-peaked $ ^{56}$Ni distribution powers both light curve peaks, with most of the $^{56}$Ni located in the center. Based on the large ejecta mass implied by the late-time main peak, the progenitor star of PTF11mnb could have been a massive, possibly single WR star, which was entirely stripped of its hydrogen and helium envelope. 

However, a hybrid model with a magnetar powering the main peak, and a normal SN explosion powering the first peak (as suggested for SN~2005bf) cannot be excluded. In this case the ejecta mass would be significantly lower. 

The early $g-$band light curve emission suggests the presence of an extended envelope surrounding the progenitor of PTF11mnb, as in the case of other massive SN~Ic progenitors.

\label{sec:conclusions}

\begin{acknowledgements}
We gratefully acknowledge the support from the Knut and Alice
Wallenberg Foundation. Based on observations obtained with the Samuel Oschin 48-inch Telescope and the 60-inch Telescope at the Palomar Observatory as part of the intermediate Palomar Transient Factory (iPTF) project, a
scientific collaboration among the California Institute of Technology,
Los Alamos National Laboratory, the University of Wisconsin,
Milwaukee, the Oskar Klein Center, the Weizmann Institute of Science,
the TANGO Program of the University System of Taiwan, and the Kavli
Institute for the Physics and Mathematics of the Universe. 
The Oskar Klein Centre is funded by the Swedish Research Council. 
We thank I.  Arcavi, A. Horesh, T. Kupfer, D. Levitan, T. Matheson, G. Smadja, S. Tendulkar, and D. Xu for their help with the spectral observations and their early reductions. 
We thank Y. Cao, J. Surace,  R. Laher, F. Masci, U. Rebbapragada, and P. Wo\'zniak for their contribution to the iPTF project. 
This work is partly based on observations made with DOLoRes@TNG. 
A.G.-Y. is supported by the EU/FP7 via ERC grant No.
725161, the Quantum Universe I-Core program by the Israeli Committee for planning and funding, and the ISF,
Minerva and ISF grants, WIS-UK ``making connections'',
and Kimmel and YeS awards. G.S. acknowledges support from the Lyon Institute of Origins under grant ANR-10-LABX-66.

\end{acknowledgements}

\bibliographystyle{aa}

\onecolumn

\begin{deluxetable}{|cc|cc|cc|cc|}
\tabletypesize{\scriptsize}
\tablewidth{0pt}
\tablecaption{Optical photometry of PTF11mnb.\label{tab:phot}}
\tablehead{
\colhead{JD-2,455,000}&
\colhead{$B$}&
\colhead{JD-2,455,000}&
\colhead{$g$}&
\colhead{JD-2,455,000}&
\colhead{$r$}&
\colhead{JD-2,455,000}&
\colhead{$i$}\\
\colhead{(days)}&
\colhead{(mag)}&
\colhead{(days)}&
\colhead{(mag)}&
\colhead{(days)}&
\colhead{(mag)}&
\colhead{(days)}&
\colhead{(mag)}}
\startdata
842.803 & 19.096(0.060) & 804.857 & 21.071(0.309)   & 809.882 & 21.146(0.195) &  842.801 & 18.852(0.049) \\ 
848.672 & 18.862(0.064) & 806.881 & 20.987(0.045)   & 811.870 & 20.862(0.093) &  848.671 & 18.745(0.046) \\ 
851.679 & 19.037(0.044) & 807.880 & 20.948(0.041)   & 812.845 & 20.479(0.056) &  851.677 & 18.669(0.004) \\ 
854.666 & 19.064(0.011) & 808.906 & 20.761(0.090)   & 813.808 & 20.444(0.165) &  854.663 & 18.644(0.022) \\ 
857.694 & 19.081(0.014) & 825.748 & 19.904(0.069)   & 821.749 & 20.017(0.026) &  857.692 & 18.690(0.041) \\ 
858.656 & 19.210(0.019) & 826.915 & 19.910(0.014)   & 822.753 & 19.889(0.073) &  858.654 & 18.677(0.015) \\ 
862.656 & 19.405(0.030) & 827.939 & 19.816(0.042)   & 823.750 & 19.772(0.056) &  862.653 & 18.721(0.024) \\ 
866.667 & 19.514(0.029) & 828.946 & 19.916(0.031)   & 824.787 & 19.870(0.041) &  866.664 & 18.821(0.029) \\ 
869.633 & 19.667(0.124) & 829.925 & 19.857(0.022)   & 824.896 & 19.645(0.017) &  869.630 & 18.855(0.009) \\ 
916.694 & 20.886(0.142) & 830.956 & 19.846(0.047)   & 826.985 & 19.686(0.026) &  891.828 & 19.263(0.074) \\ 
922.673 & 20.847(0.078) & 831.958 & 19.839(0.006)   & 828.946 & 19.554(0.016) &  903.665 & 19.507(0.083) \\ 
         &              & 833.768 & 19.828(0.017)   & 838.672 & 19.199(0.060) &  906.794 & 19.624(0.087) \\ 
         &              & 834.768 & 19.701(0.020)   & 842.711 & 19.082(0.009) &  916.691 & 19.825(0.078) \\ 
         &              & 835.800 & 19.551(0.015)   & 849.636 & 18.812(0.104) &  922.670 & 19.680(0.053) \\ 
         &              & 836.796 & 19.481(0.026)   & 850.632 & 18.742(0.028) &  930.653 & 19.700(0.086) \\ 
         &              & 837.819 & 19.390(0.051)   & 851.631 & 18.797(0.007) &  944.688 & 19.866(0.095) \\ 
         &              & 842.805 & 18.972(0.041)   & 852.749 & 18.789(0.023) &  945.623 & 19.925(0.058) \\ 
         &              & 848.674 & 18.821(0.054)   & 853.737 & 18.764(0.001) &  952.637 & 20.148(0.103) \\ 
         &              & 851.680 & 18.830(0.018)   & 854.716 & 18.791(0.033) &          &               \\
         &              & 854.668 & 18.901(0.000)   & 862.655 & 18.694(0.016) &          &               \\
         &              & 855.639 & 18.775(0.002)   & 866.665 & 18.824(0.027) &          &               \\
         &              & 856.728 & 18.775(0.011)   & 869.632 & 18.867(0.019) &          &               \\
         &              & 857.696 & 18.887(0.008)   & 875.588 & 18.785(0.086) &          &               \\
         &              & 857.729 & 18.819(0.020)   & 891.830 & 19.364(0.146) &          &               \\
         &              & 858.658 & 18.910(0.008)   & 903.667 & 19.552(0.055) &          &               \\
         &              & 858.732 & 18.850(0.027)   & 916.692 & 19.815(0.072) &          &               \\
         &              & 859.737 & 18.923(0.038)   & 922.672 & 19.827(0.048) &          &               \\
         &              & 861.654 & 18.921(0.005)   & 923.610 & 19.700(0.055) &          &               \\
         &              & 862.658 & 19.113(0.017)   & 930.654 & 19.652(0.059) &          &               \\
         &              & 862.715 & 19.058(0.029)   & 944.690 & 19.887(0.103) &          &               \\
         &              & 863.714 & 19.113(0.033)   & 952.638 & 20.031(0.099) &          &               \\
         &              & 864.729 & 19.164(0.023)   &         &               &          &               \\
         &              & 865.792 & 19.177(0.027)   &         &               &          &               \\
         &              & 866.668 & 19.296(0.023)   &         &               &          &               \\
         &              & 869.638 & 19.420(0.064)   &         &               &          &               \\
         &              & 903.670 & 20.684(0.130)   &         &               &          &               \\
         &              & 916.695 & 20.759(0.099)   &         &               &          &               \\
         &              & 923.612 & 20.618(0.101)   &         &               &          &               \\
         &              & 930.658 & 20.625(0.163)   &         &               &          &               \\
         &              & 944.693 & 20.672(0.171)   &         &               &          &               \\
  \enddata
\end{deluxetable}

\begin{deluxetable}{cccccc}
\tabletypesize{\scriptsize}
\tablewidth{0pt}
\tablecaption{Optical spectroscopy of PTF11mnb and its host-galaxy \label{tab:spectra}}
\tablehead{
\colhead{Date (UT)}&
\colhead{JD-2,455,000}&
\colhead{Phase\tablenotemark{a}}&
\colhead{Telescope}&
\colhead{Instrument}&
\colhead{Range}\\
\colhead{}&
\colhead{(days)}&
\colhead{(days)}&
\colhead{}&
\colhead{}&
\colhead{(\AA)}}
\startdata
{\bf SN spectra}& & & & & \\


 07 Oct. 2011  & 841.98  &  +35.5 & UH88   &   SNIFS   & 3301$-$9701 \\
  30 Oct .2011  & 864.78  &  +57.0 & P200   &   DBSP    & 3002$-$10293 \\
   26 Nov. 2011  & 891.92  &  +82.6 & Keck I   &   LRIS    & 3001$-$10297 \\
    21 Dec. 2011  & 916.68  &  +105.9 & P200   &   DBSP    & 3001$-$10296 \\
     27 Jan. 2012  & 953.50  &  +140.7 & KPNO4m   &   RC Spec    & 3575$-$8138 \\
     \hline
     {\bf Host-galaxy spectrum}& & & & & \\
      02 Sep. 2016  & \ldots  &  \ldots & TNG   &   DOLORES    & 3189$-$10352 \\    
\enddata
\tablenotetext{a}{Rest frame days from explosion.}
\end{deluxetable}

\begin{table}
\centering
\caption{\label{tab:propgal}Properties of PTF11mnb and its host galaxy SDSS J003413.34$+$024832.9}
\vskip 0.2cm
\scriptsize
\begin{tabular}{llc}
\hline\hline
     Parameters&Value&Ref\\
     \hline
     {\bf SDSS J003413.34$+$024832.9:}&&\\
     \\
     Position & $\alpha_{\rm J2000} = 00^{\rm h} 34^{\rm m} 13\fs34$&Sect. \ref{sec:hg}\\
              & $\delta_{\rm J2000} = +02\degr 48\arcmin 32\farcs9$&  \\
     \\
     Abs. magnitude& $M_{g}=-17.8$ mag&Sect. \ref{sec:hg}\\
         & $M_{r}=-18.0$ mag& \\
         & $M_{i}=-18.3$ mag& \\
     \\
     Redshift& $z=0.0603\pm0.0001$ &Sect. \ref{sec:hg}\\
     \\
     \\
     Distance& $D=268.5$\,Mpc&Sect. \ref{sec:hg}\\
     \\
     Distance modulus& $\mu = 37.14$\,mag&Sect. \ref{sec:hg}\\
     \\
     \\
     Metallicity of the host & log(O/H)$+$12$\,$=8.29$\,\pm$0.20&Sect. \ref{sec:hg}\\
     at the SN location (O3N2)& & \\
     \\
     {\bf PTF11mnb:}&&\\

  Position & $\alpha_{\rm J2000} = 00^{\rm h} 34^{\rm m} 13\fs25$&Sect. \ref{sec:obs}\\
              & $\delta_{\rm J2000} = +02\degr 48\arcmin 31\farcs4$&  \\
     \\
     Explosion epoch & JD 2455804.341$\pm$0.516 & Sect. \ref{sec:obs}\\
     \\
     Total reddening toward SN: & $E(B-V)$\,$=0.016$\,mag&Sect. \ref{sec:hg}\\ 
     
 \\         Phase of the main peak, in& $t_{B}^{max}~=$~41.9~d &Sect. \ref{sec:lc}\\ 
rest-frame days since explosion & $t_{g}^{max}~=$~46.3~d &\\ 
 & $t_{r}^{max}~=$~52.2~d &\\ 
 & $t_{i}^{max}~=$~48.1~d &\\ 

  \\         Peak abs. magnitude& $B_{max}~=$~$-$18.25~mag &Sect. \ref{sec:lc}\\ 
& $g_{max}~~=$~$-$18.37~mag &\\ 
 & $r_{max}~~=$~$-$18.45~mag &\\ 
 & $i_{max}~~=$~$-$18.52~mag &\\ 
 
  \\         $\Delta m_{15}$& $\Delta {B}_{15}~=$~0.48~mag &Sect. \ref{sec:lc}\\ 
& $\Delta {g}_{15}~=$~0.53~mag &\\ 
 & $\Delta {r}_{15}~=$~0.16~mag &\\ 
 & $\Delta {i}_{15}~=$~0.24~mag &\\ 
    
     \hline
\end{tabular}
\end{table}

\end{document}